\documentclass{aa}
\usepackage{natbib,amsmath,amssymb,graphicx,graphics}
\usepackage{txfonts}
\usepackage{subfigure}

\def\bp{\object{$\beta$\,Pictoris}}

\begin{document}
\title{VLT/NACO coronagraphic observations of fine structures in the disk
  of $\beta$ Pictoris}
\titlerunning{Fine structures in the disk of $\beta$ Pictoris}

\author{
A. Boccaletti\inst{1} \and 
J.-C.  Augereau\inst{2} \and
P. Baudoz\inst{1} \and
E. Pantin\inst{4} \and
A.-M. Lagrange\inst{2}
}
\institute{LESIA, Observatoire de Paris-Meudon F-92195, Meudon, France \\
  \email{anthony.boccaletti@obspm.fr, pierre.baudoz@obspm.fr} 
  \and Laboratoire d'Astrophysique de Grenoble, Universit\'e Joseph Fourier, CNRS, UMR 5571, Grenoble, France \\
  \email{augereau@obs.ujf-grenoble.fr} 
  \and Service d'Astrophysique, CEA Saclay, F-91191 Gif-sur-Yvette, France \\
  \email{epantin@cea.fr}}

\offprints{A. Boccaletti}
\date{Received ; accepted }

\abstract
{}
{We present ground-based observations of the disk around the A-type
  star $\beta$ Pictoris to obtain scattered light images
  at the highest angular resolution ($60$\,mas, equivalent to about
  $1$\,AU at the distance of the star) and the highest contrast in the
  very close environment of the star. The purpose of this program is
  to perform a close inspection of the inner disk morphology.}
{Images were collected with NACO, the AO-assisted near-IR instrument
  on the VLT (ESO) which includes two types of coronagraphs: classical
  Lyot masks and phase masks. In this program we took advantage of
  both types of coronagraphs in two spectral bands, H-band for the
  Lyot mask and Ks-band for the phase mask. The Lyot mask blocks
  a large central region around the star (radius $<0.35$") but allows
  deep integrations and hence good signal-to-noise ratio at large
  distances, while the phase mask allows imaging at very close
  separation (down to $\sim0.15$" in theory) but conversely is more
  sensitive to residual aberrations. In addition, we simulated an extended object to understand the limitations in deconvolution of coronagraphic images. }
{The reduced coronagraphic images allow us to carefully measure the
  structures of the debris disk
  and reveal a number of asymmetries of which some were not reported
  before (position, elevation and thickness of the warp). Our analysis also demonstrates
  the advantage of the phase mask coronagraph to explore the very
  close environment of stars. In this program, the circumstellar
  material is visible as close as $0.7$" ($13.5\,$AU) owing to the
  phase mask while the Lyot mask generates artifacts which hamper the
  detection of the dust at separations closer than $1.2$"
  ($23.2\,$AU). The point source detection limit is compared to recently published observations of a planet candidate.
  Finally, the simulations show that deconvolution of coronagraphic data may indeed produce artificial patterns within the image of a disk.
  }
{} 

  \maketitle

\section{Introduction}
The young ($<$20\,Myr) and nearby A-type star \bp\ has been the subject of
many investigations since the imaging by \citet{smith84} of the
first circumstellar disk around a star. Over the last 25 years, the
edge-on dusty disk was observed at many wavelengths, in scattered light
and in thermal emission, from space and from the ground with the
prime goal to understand the planet formation process, and ultimately to
find evidence for planetary objects.

\begin{figure*}
  \centerline{
    \includegraphics[width=9cm]{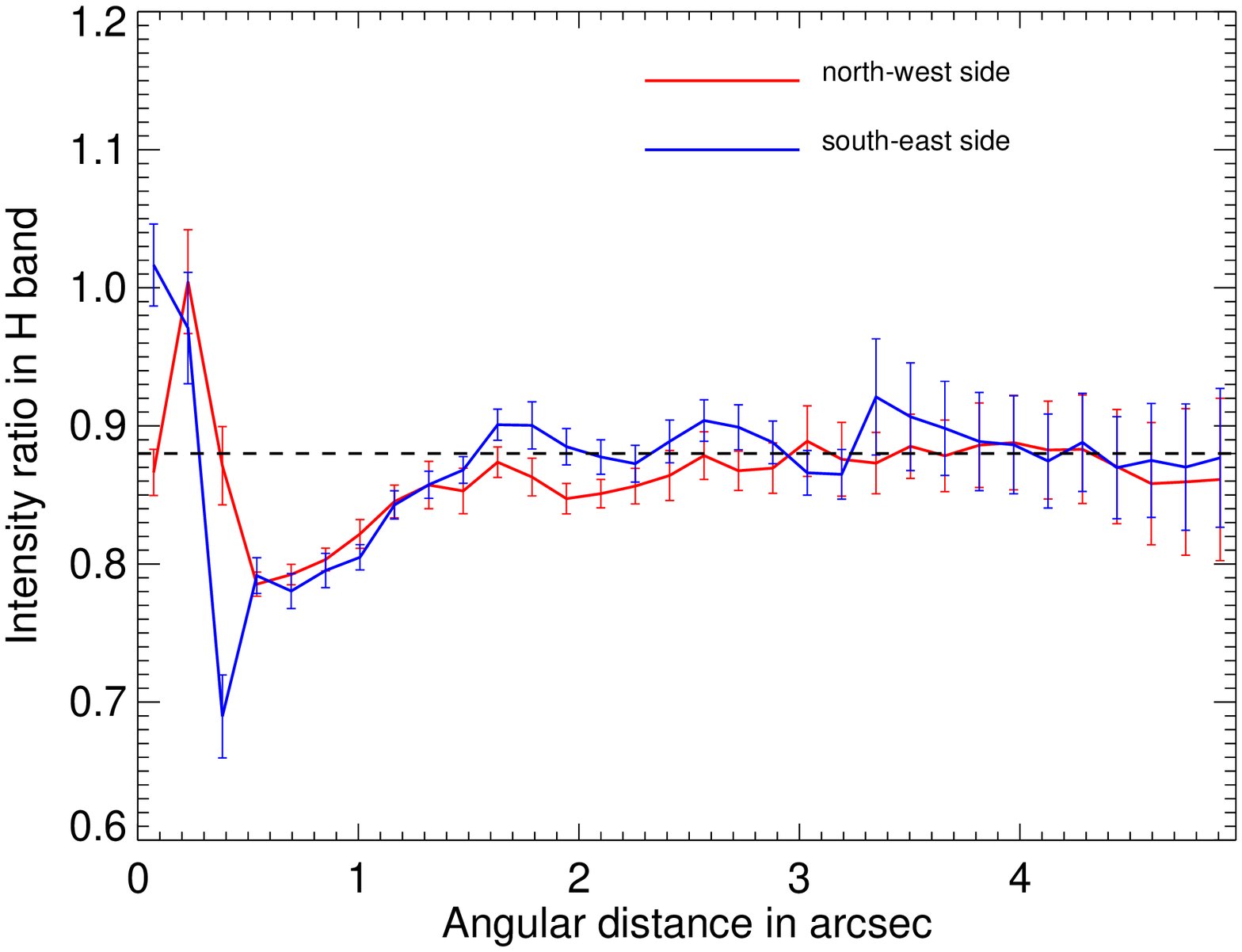}
    \includegraphics[width=9cm]{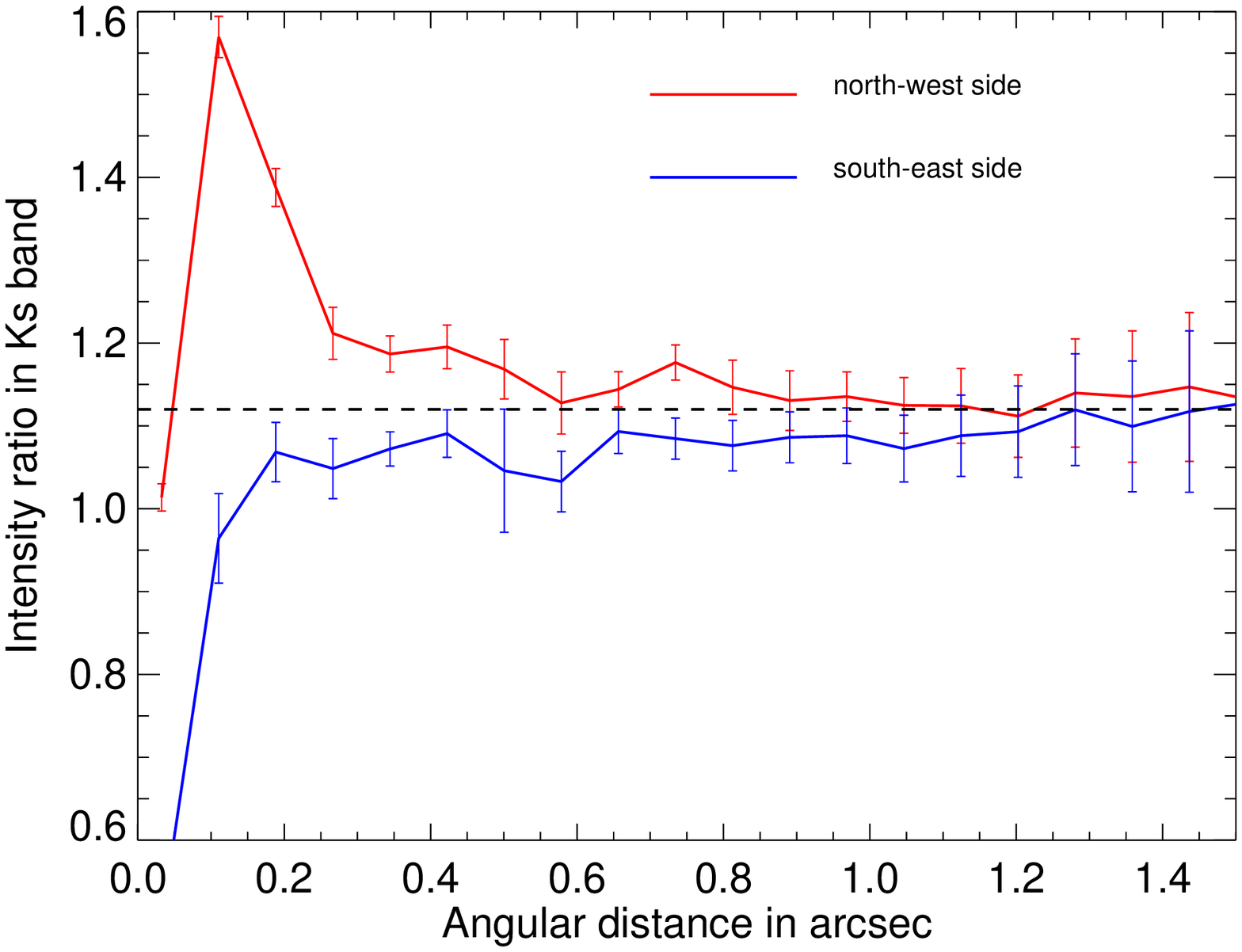}}
  \caption[]{Target to reference flux ratios as a function of the
    angular separation to the star for the Lyot image (left panel) and
    the FQPM image (right panel). The red and blue curves are the average values measured in angular sectors of 30$^\circ$
    and orthogonally to the disk midplane (north-west and south-east directions). The vertical bars correspond to the 1$\sigma$
    dispersion in these sectors. The adopted scaling factors  are shown as horizontal dashed lines.
    }
  \label{fig:ratio}
\end{figure*}

The discoverers, \citet{smith84}, resolved the disk in the visible at
projected separations larger than 5.2" ($\sim$100\,AU, adopting the
Hipparcos distance of 19.3\,pc), and, extrapolating the observed surface brightness profile to shorter
distances, they proposed that the disk should be optically thick below
15\,AU. About ten years latter, \citet{kalas95} observed the disk down
to 2.8" (54\,AU) and identified morphological and brightness
asymmetries between the north-east and south-west extensions of the
edge-on disk. The WFPC2 instrument onboard the HST was used to observe
the disk as close as 1.5" (29\,AU) and confirms the asymmetrical
structure. 
Mid-infrared imaging \citep{lagage94} of the innermost regions of the disk
(0-80\,AU) showed that the inner disk ($r<$25\,AU) is depleted by a factor 
of $\sim 100$ with respect to the regions of peak density ($r \approx 80$\,AU).
Based on new datasets, this depletion was confirmed by \citet{pantin97} who proposed 
a two-component model to interpret both the visible/near-infrared and mid-infrared
data. 
From modeling, \citet{burrows95} derived the presence of a relatively
clear zone within 40\,AU. In addition, they revealed a peculiar
symmetrical structure, the so-called warp, at a distance $<$70 AU
which should have disappeared in about 1\,Myr.  They interpreted this
feature as the gravitational signature of a planetary object which
would also cause the clearing of the inner disk regions.  With the
improvement of Adaptive Optics system (AO) on large ground-based
telescopes, the warp was confirmed by \citet{mouillet97a} in the
near-IR using the 3.6\,m telescope at La Silla (ESO).  To account for
the presence of the warp \citet{mouillet97b} developed a model
which concluded the presence of a planet on an inclined orbit
(3-5$^\circ$) and located in the range 1-20\,AU from the star.  This
warp was also observed in detail with the HST/STIS instrument at
visible wavelengths \citep{heap00}. The small Lyot mask and the
quality of the PSF subtraction process allow the detection of the
disk as close to the star as 0.75". According to \citet{heap00}, the
perturbing planet may have a mass of a few times that of Jupiter if
located closer than 20\,AU. The most recent analysis
\citep{Freistetter07} based on numerical simulations constrained by
the presence of planetesimal belts inferred from the observed temporal
variations of metallic lines in the \bp\ spectrum suggests a planet of
2\,M$_J$ at 12\,AU. Note that high resolution spectra were also used
 to search for radial velocity variations of \bp \citep{Galland06}. The
available data allowed them to detect pulsations (also seen in photometry)
and to constrain the presence of planets with small periods (e.g. a
few days). Longer periods planets are still poorly constrained with
radial velocity data, with for example, at 1\,AU, a limit of 9\,M$_J$.
While this paper was being reviewed, a candidate planet was discovered by \citet{Lagrange08} taking advantage of a reduced contrast in L' band images. With an inferred mass of 8\,M$_J$, at a projected distance of 8\,AU it is significantly more massive than the \citet{Freistetter07}  predictions although more measurements would be needed to definitely rule on the mass of this candidate planet and confirm the companionship.

A detailled dynamical model of the \bp\ dust disk was developed by
\citet{augereau01} who simulate a disk of planetesimals and dust
particles pertubed by a giant planet on an inclined orbit with
respect to the disk midplane. The \citet{kalas95} large scale
asymmetries (at hundreds of AUs) and the warp observed in scattered
light are well reproduced assuming a planet located at 10\,AU about
$10^{3}$ times less massive than the central star. The inferred surface
density for the disk corresponds to a broad annulus peaked around
80--110\,AU and smoothly declining inward within about 70\,AU.
Although the far-infrared thermal emission is correcty predicted by
this model, it is shown that most of the mid-infrared emission is very
likely produced by an additional hot population of small grains close to the star
and not modeled.  Recent high-resolution mid-IR images
\citep{telesco05} of the inner disk obtained with Gemini have revealed
the presence of an asymmetric structure located at about 50\,AU and
which may correspond to a clump of particles differing in temperature
and/or in size to the other particles in the disk. This observation
supports the commonly shared hypothesis that the disk is replenished
by short-lived dust originating from collisions of planetesimals but
the mechanisms that can produce a clump of that size are still
unclear.

The latest images of the \bp\ disk were obtained by \citet{golimowski06} in the
visible with ACS on the HST. Despite the very high sensitivity of the
ACS, the large opaque Lyot mask does not allow them to probe the disk
morphology at an angular separation less than about 1.5" to the star.  Using a
specific data reduction process to emphasize high spatial frequency
structures in the disk, they suggest that the warp is in fact the
result of a blend of a main disk with a much fainter, inclined
secondary disk which can be separated from the main one from 80\,AU up
to about 150\,AU in ACS images. However, the method used to separate
the two disks is arguable (see Sec.\,\ref{sec:warp} for a more
detailled discussion) and it is therefore not clear if this secondary
disk is a fully distinct component or originates in the main, twisted
disk as supported by the work of \citet{mouillet97b} and
\citet{augereau01}.  \citet{golimowski06} also showed an
interesting color dependence of the disk in scattered light from which
they inferred a typical minimum grain size of about $0.2\,\mu$m where
the planetesimals lie ($r<120$\,AU).

Improving the angular resolution in the first few tens of AU from \bp\
is now clearly mandatory to access the planet forming regions.
Optimal conditions are met either with large ground-based telescopes
(providing the wavefront is corrected with high-order AO systems) or
from space with the HST. In this paper, we present new
coronagraphic images obtained with an 8\,m telescope at the European
Southern Observatory in the near-IR. Although the sensitivity is not
sufficient to detect giant planets, the coronagraph we used is
potentially sensitive to smaller separations than any previous
observations. In section \ref{sec:obs} we describe the observations we
carried out and the data reduction process. The analysis of the disk
in terms of morphology, surface brightness profile, vertical flux and
position of the midplane is presented in section
\ref{sec:analysis}. We then discuss the hypothesis of a secondary disk
(section \ref{sec:warp}) and the contamination by PSF structures in
coronagraphic images (section \ref{sec:deconv}).

\begin{figure*}
  \centerline{
    \includegraphics[width=9cm]{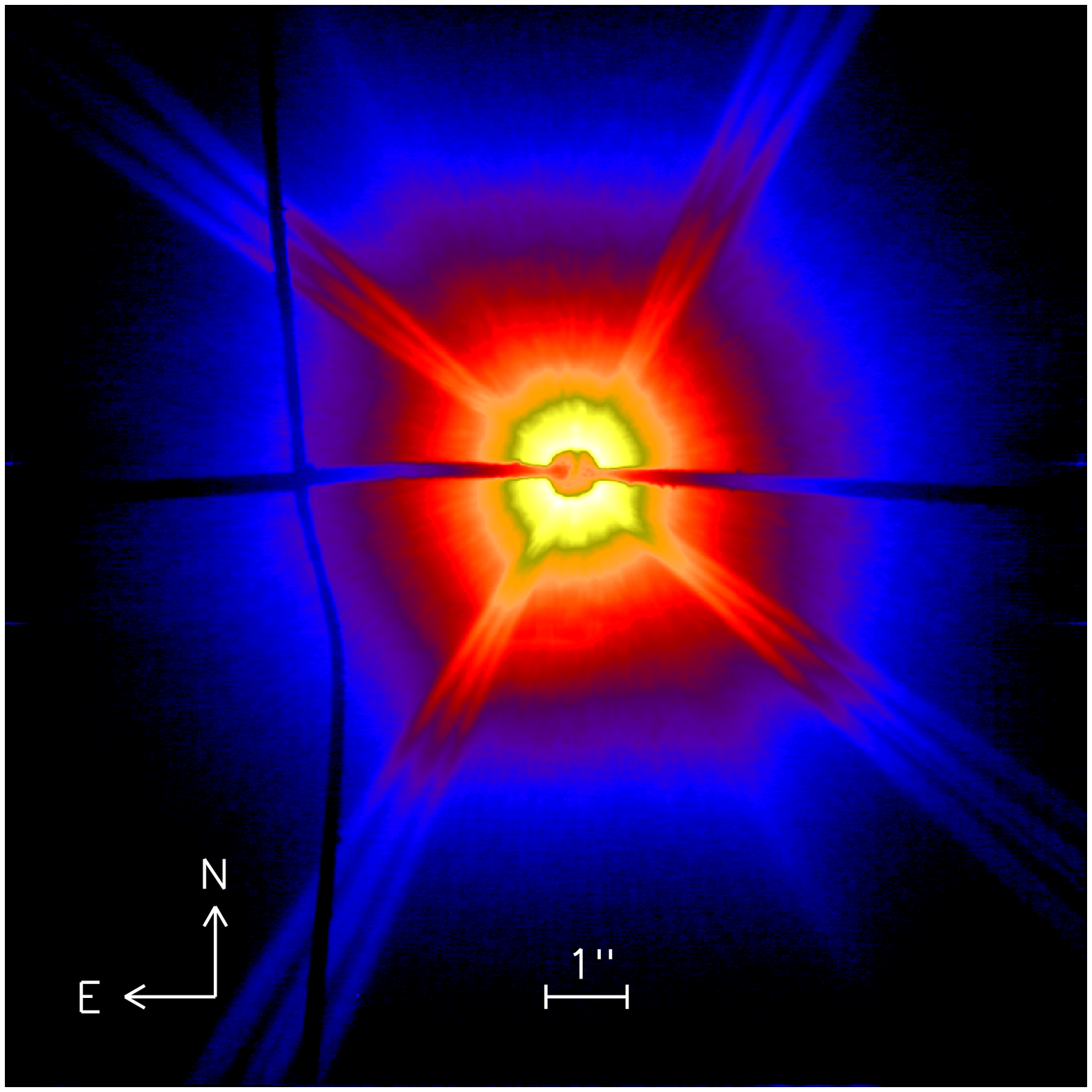}
    \includegraphics[width=9cm]{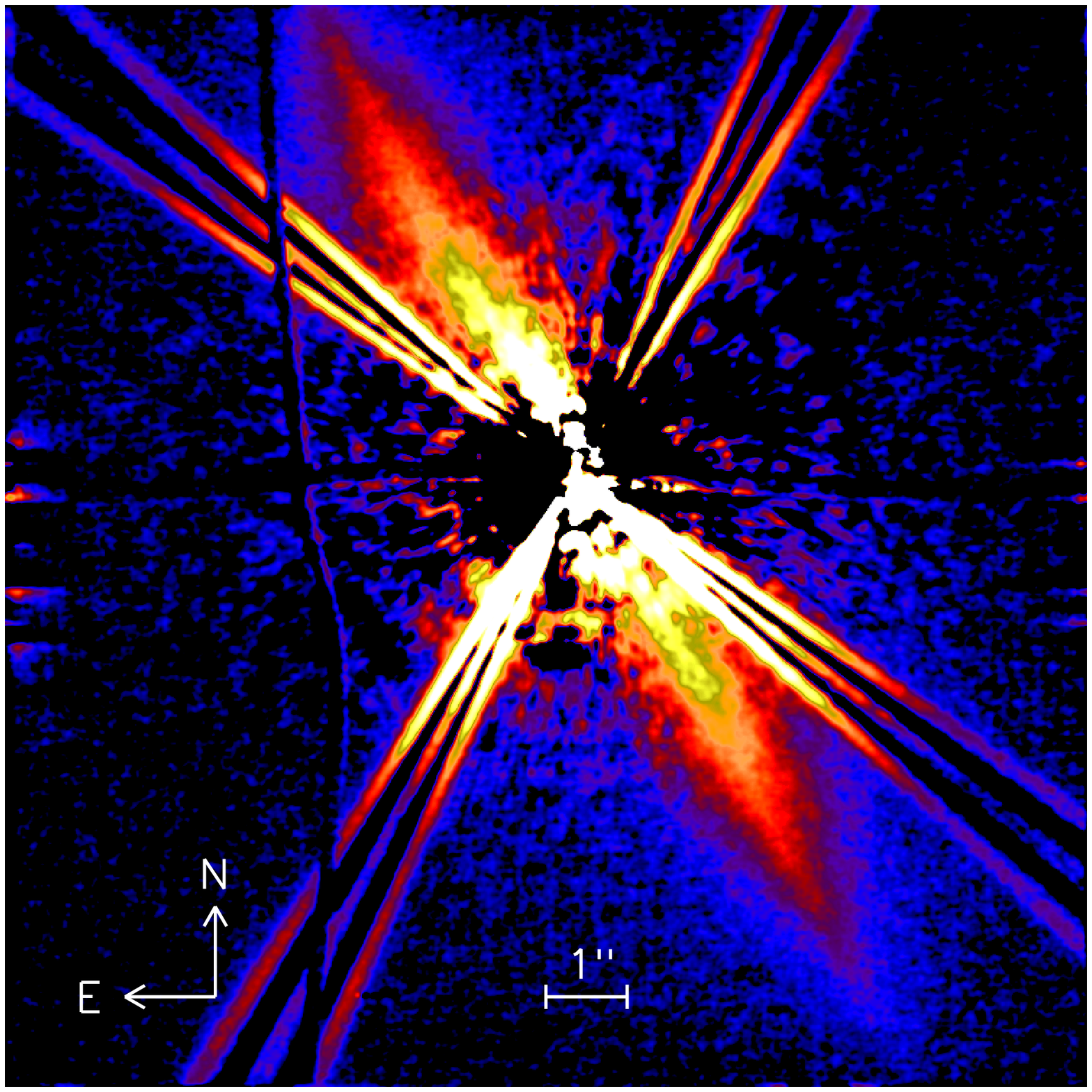}}
  \caption[]{Images of \bp\ observed with the Lyot coronagraphic mask
    in the H band. A raw image of the star with the 0.7" Lyot mask is displayed to the left while the picture to the right reveals the disk after subtraction of the reference star. The
    field of view is 13.3".  The diffraction by the spiders are
    localized in 3 spikes because the sky is calibrated twice during the
    observation while the pupil rotates. 
    }
  \label{fig:lyotraw}
\end{figure*}
\begin{figure*}
  \centerline{
    \includegraphics[width=9cm]{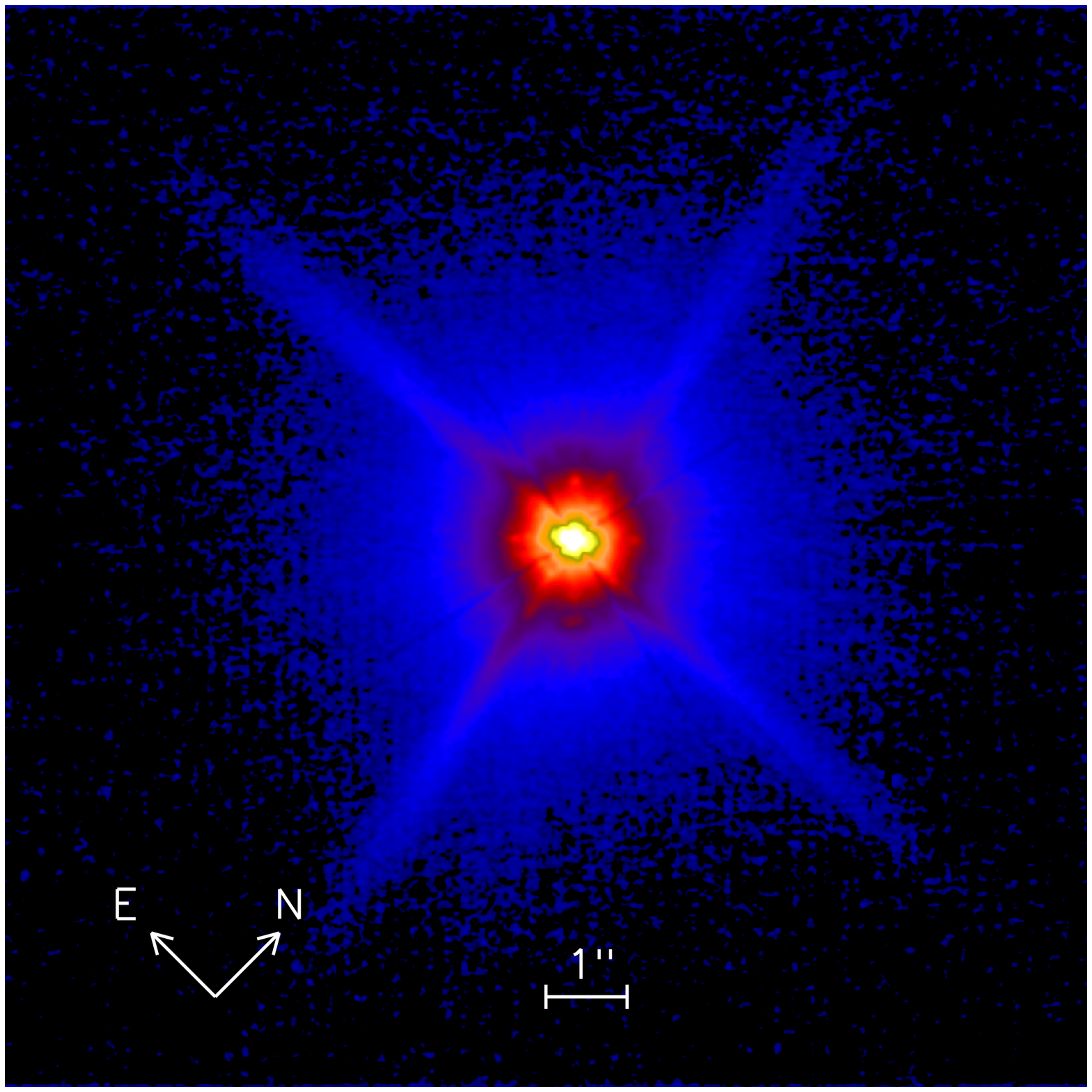}
    \includegraphics[width=9cm]{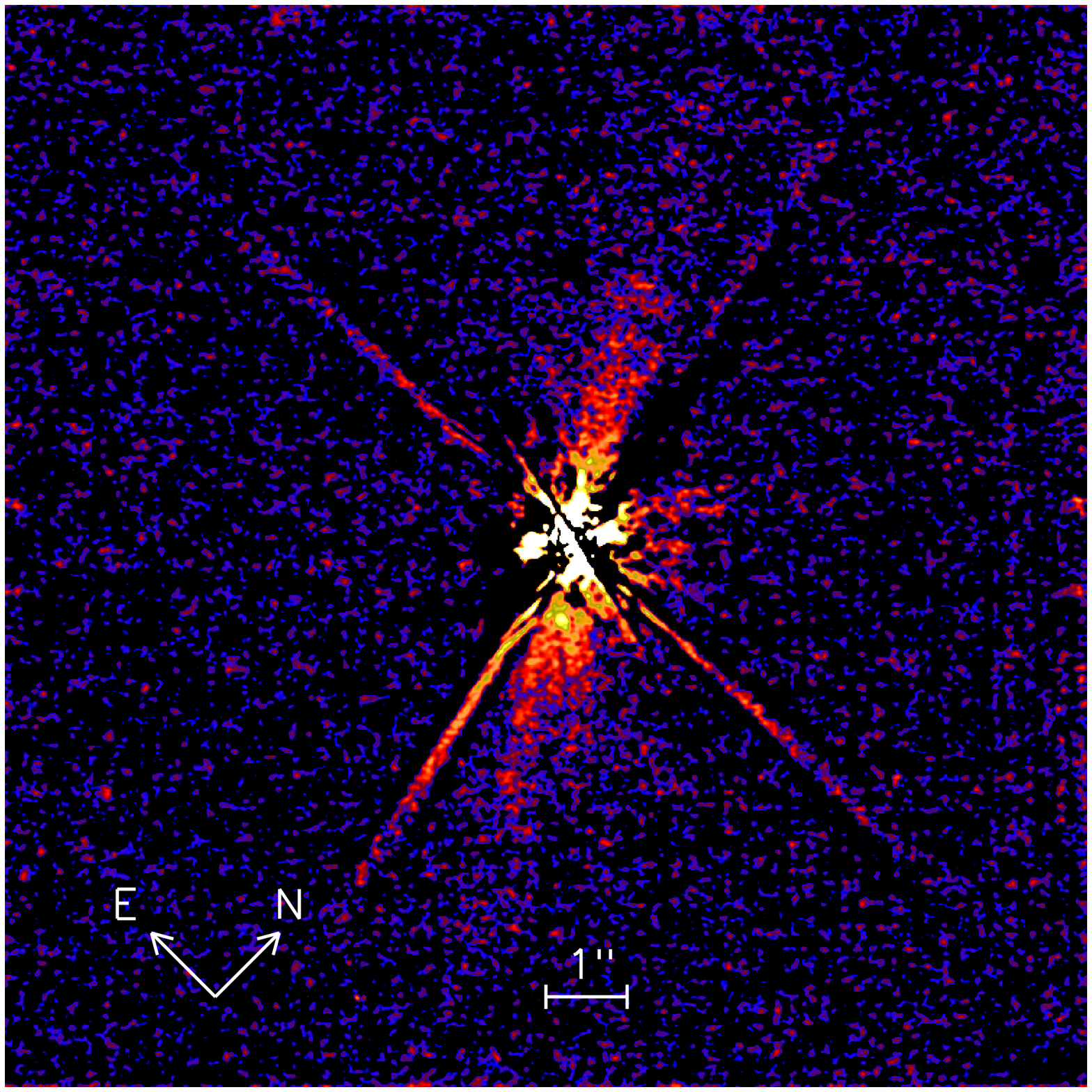}}
  \caption[]{Images of \bp\ observed with the FQPM in the K$_s$
    filter (left) and after subtraction of the reference star
    (right). The field of view is 13.3". Here, the sky is calibrated once
    at the end of the observation.
    }
  \label{fig:4qraw}
\end{figure*}

\begin{figure*}
  \centerline{\includegraphics[width=9cm]{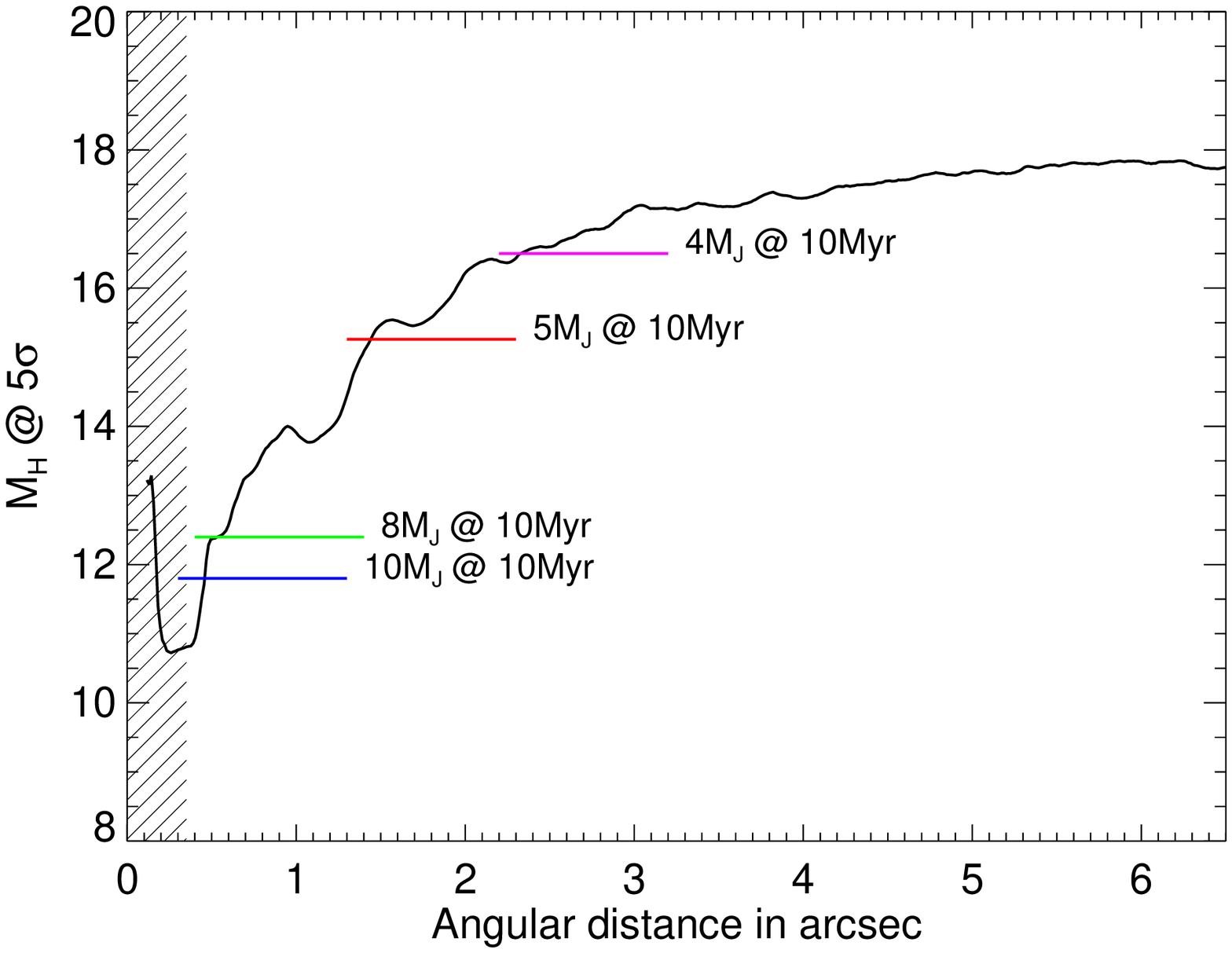}
    \includegraphics[width=9cm]{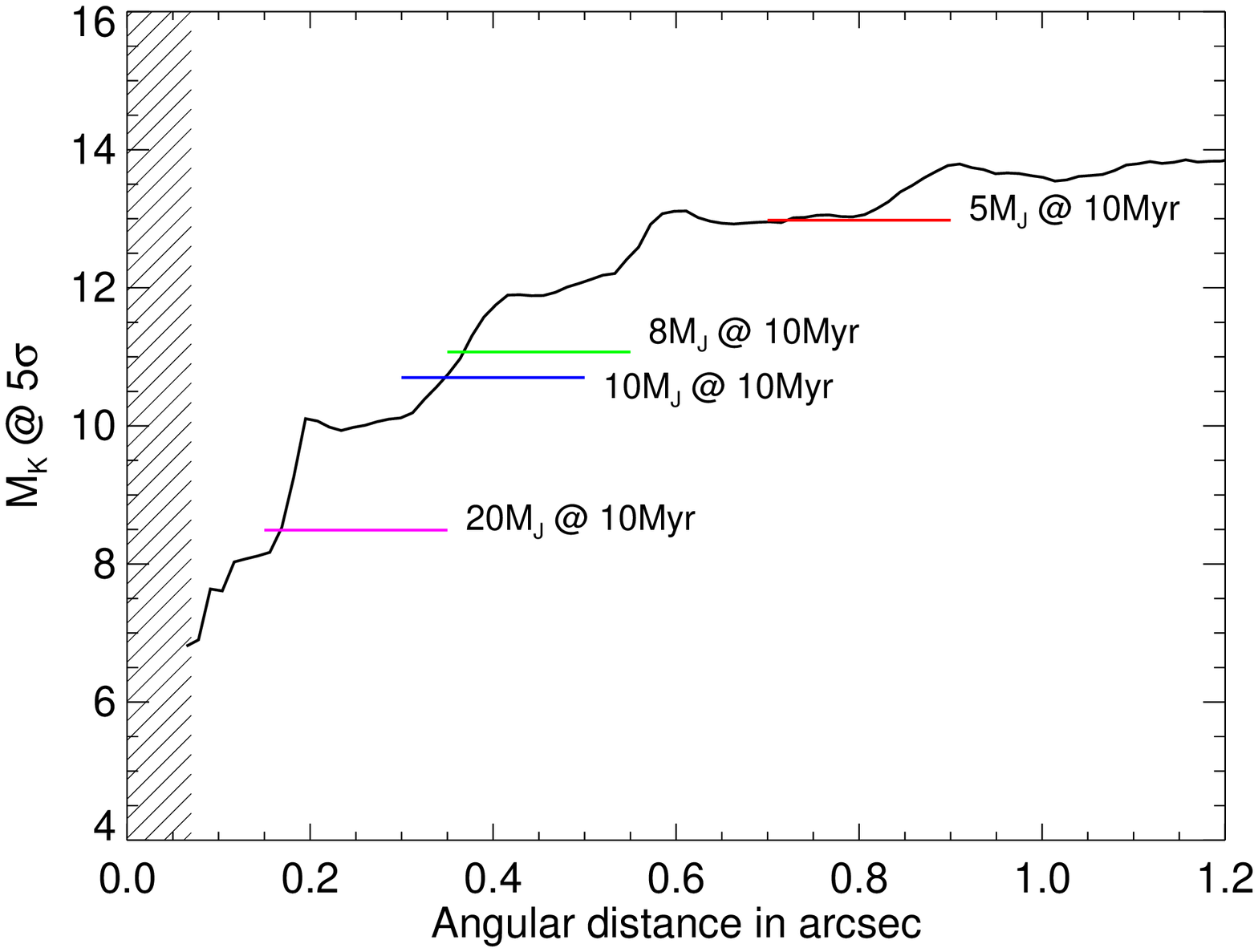}}
  \caption[]{
   Point source detection limits (expressed in absolute magnitudes) in the H and Ks bands
    as measured respectively on the Lyot (left panel) and FQPM images
    (right panel) along the disk (angular sector of $\pm 8^\circ$).
    The absolute magnitudes of some planetary
    masses are overplotted for 10\,Myr (derived from the DUSTY
    models of \citealt{Chabrier00}). The 8\,M$_{\mathrm{J}}$ limit is reached at 0.55" and 0.35" for the H and Ks bands.  
    The dashed areas have transmission lower than 50\% due to the coronagraphs (inner working angle). 100\% transmission is reached at 0.6" and 0.17" respectively for the Lyot and the FQPM. }
  \label{fig:detectionlimit}
\end{figure*}

\begin{figure}
  \centerline{\includegraphics[width=9cm]{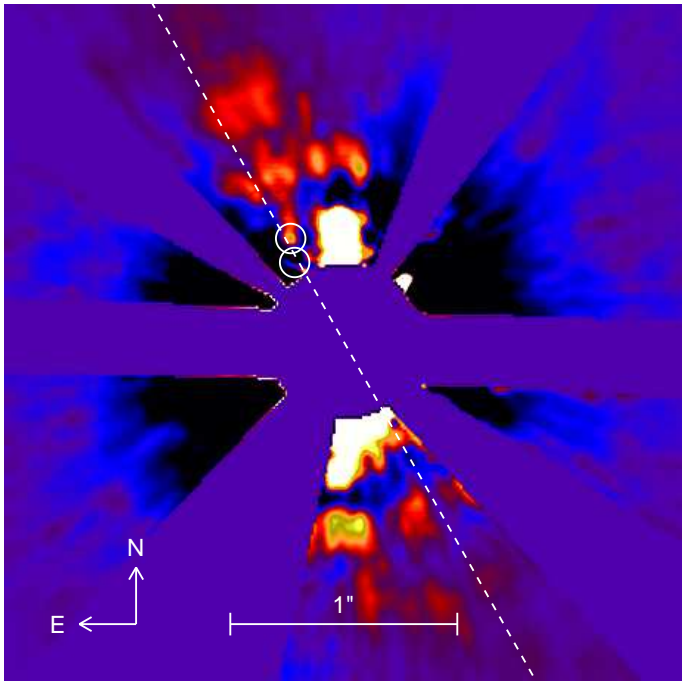}}
  \caption[]{Magnified image of the central 3" area of Fig. \ref{fig:lyotraw} (right). The lower circle shows the expected position of the candidate planet found by \citet{Lagrange08} and the upper circle indicates the suspicious point-like pattern described in sec. \ref{sec:pointsource}. Circles are twice the FWHMs. The dotted line shows the disk midplane at PA=29.5$^{\circ}$ (see section \ref{sec:analysis}).
  }
  \label{fig:circles}
\end{figure}

\section{Observations and data reduction}\label{sec:obs}
\subsection{Observing strategy}
Coronagraphic observations of \bp\ were obtained on Nov. 5th, 2004 at
ESO with NACO \citep{rousset03} the IR AO system of the VLT. \bp\
(V=3.86, H=3.54, K=3.53) was observed with 2 coronagraphs, a standard
0.7" diameter Lyot mask and a four quadrant phase mask (FQPM
hereafter), in the H and Ks spectral bands, respectively. Seeing
conditions and AO compensation were good but variable (Strehl ratios
$\sim$30-50\%). Although theoretical FWHMs are 42\,mas and 56\,mas in H and Ks filters,  
we measured respectively 61\,mas and 66\,mas resulting from both the actual Strehl ratios and the coronagraphic pupil stop (10\% undersized).

The principle of the FQPM is discussed in \citet{rouan00}. It uses a
$\pi$ phase shift in the focal plane resulting in a destructive
interference inside the geometric pupil of a point-like object
centered on the mask. A monochromatic FQPM operating at $2.15\,\mu$m
was implemented in NACO on August 2003. Performance assessment of the
FQPM is reported in \citet{boccaletti04a}. This coronagraph has no
obscuration in the focal plane and therefore is able to reach the
theoretical diffraction limit of the telescope. As a drawback, in
actual observations (Strehl ratio $<$50\%) the attenuation of the PSF
peak is less than for a Lyot coronagraph (a factor of $\sim10$ instead
of $\sim200$ depending on the level of the AO compensation). The
chromatism of the FQPM is not a limitation on NACO as demonstrated in
\citet{boccaletti04a}. Therefore, we decided to take advantage of both
coronagraphs to get (i) a good signal to noise in the disk at large
angular separations using the Lyot coronagraph and (ii) to
characterize the circumstellar dust at very close separations with the
FQPM.

The reference star HD\,45291 (V=5.97, H=3.84, K=3.77), which was
observed right after the target star, was chosen to have the same
parallactic angle as \bp\ in order to preserve the same pupil
orientation. This way, the diffraction spikes of the reference star
have about the same orientation as those of \bp\ with respect to the
field of view. The reference star has the same declination but a
difference in right ascension which depends on the integration time
(36\,minutes here). We observed \bp\ for a total of 800\,s in the H
band (Lyot coronagraph), and 700\,s with the FQPM in the Ks band (but
with a transmission of only 10\% to avoid saturation of individual
frames), and similarly for the reference star. In addition, the
field of view was rotated by 45$^\circ$ when the FQPM is used to place the disk 
in a different direction than that of the phase transitions of the phase mask. The pixel sampling is
13\,mas/pixel.

\begin{figure*}
\centerline{\includegraphics[trim= 5mm 5mm 5mm 5mm ,clip,width=15cm]{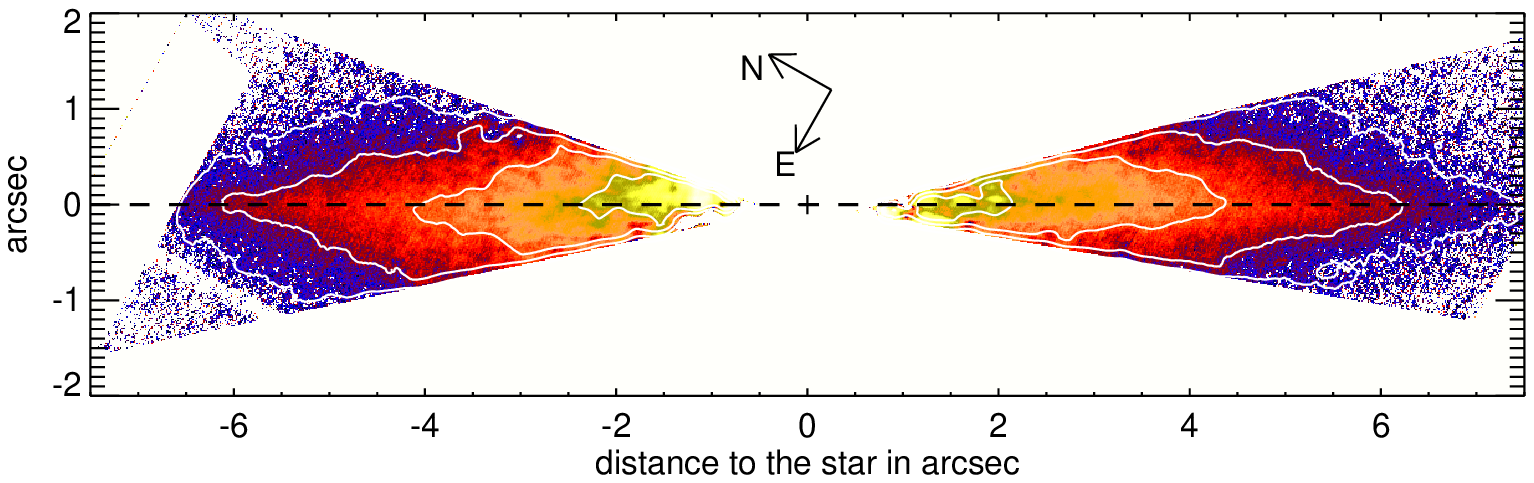}}
\centerline{\includegraphics[trim= 5mm 5mm 5mm 5mm ,clip,width=15cm]{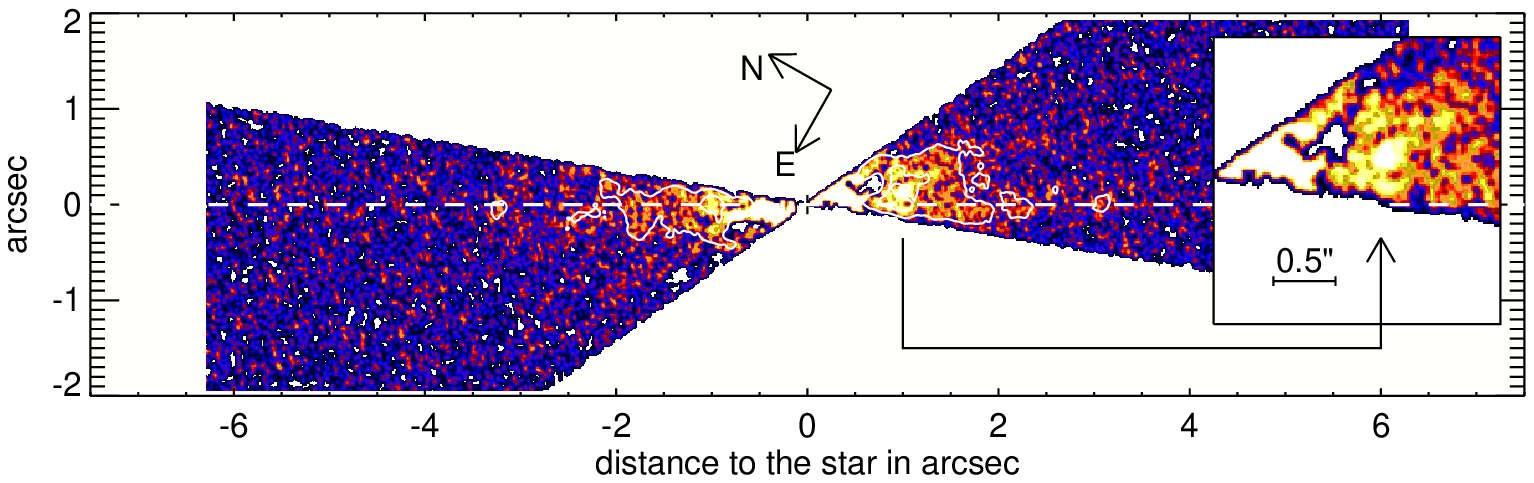}} 
\centerline{\includegraphics[trim= 5mm 5mm 5mm 5mm ,clip,width=15cm]{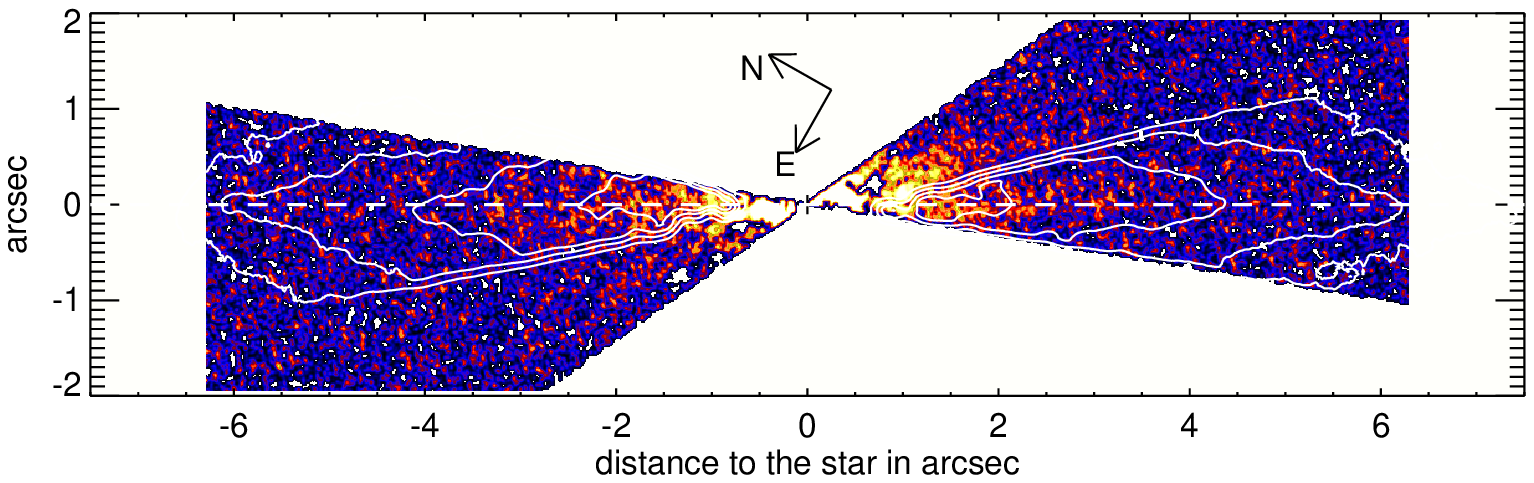}}
\caption[]{Comparison of \bp\ images and contour lines. The
  contours on the Lyot image (top) correspond to 12, 13, 14 and 15
  mag/arcsec$^2$ while on the FQPM image (middle) the lines stand for
  12 and 12.5 mag/arcsec$^2$. These contours are obviously shaped by the numerical sector mask especially in the Lyot image (top). The image at the bottom shows again the FQPM image overplotted with the contours of the Lyot image. The spine is shown as a dashed line.} 
\label{fig:horizontal}
\end{figure*}

%
\subsection{Data reduction}
\label{sec:reduction}
During the observations, the individual exposures (DIT\,=\,1\,s) are
combined in long exposure images of 100\,s each
(NDIT\,=\,100). Cosmetic reduction is applied on each image including
flat fielding, bad pixel correction and sky subtraction. Images are then re-centered and co-added
independently for the target star and the reference star.

Our procedure to remove the stellar residuals (unattenuated by the
coronagraph) using a reference star is similar for the Lyot and the
FQPM images. As described for example in \citet{augereau99} and
\citet{boccaletti03}, the image of the target is first divided by that
of the reference (re-centered at a subpixel scale) to provide a map of
the intensity ratio.  The scaling factor is measured at a radial
distance where the intensity ratio map is almost flat ($\rho=1.5-5$" on
the Lyot image and $\rho=0.2-1.5$" on the FQPM image) and is
azimuthaly averaged in regions free of circumstellar material
(perpendicularly to the disk midplane, in the north-west and south-east directions). 
The intensity ratio profiles are displayed in Fig.\,\ref{fig:ratio}.

The reference star image is then normalized to the scaling factor
and subtracted from the target star image. After this process, the Lyot
subtracted image still has a diffraction pattern
originating from the high frequency zonal polishing defects of the
primary mirror (\ref{fig:lyotraw}).  Similar patterns are observed in HST images of
\object{AU\,Mic} \citep{krist05} for instance. This feature was
estimated in several sectors around the star and an azimuthaly
averaged pattern was subtracted. However, a
residual zonal pattern remains in the subtracted image and produces a
modulation of the disk intensity in some particular
regions. Nevertheless, the disk is clearly visible all across the
field of view. The raw coronagraphic images and the subtracted images are displayed in Fig.\,\ref{fig:lyotraw} and in Fig.\,\ref{fig:4qraw}  for the Lyot and the FQPM.

For the photometric calibration, out-of-mask images of the star
were collected, but with a different set-up than coronagraphic images. 
Indeed, these calibration data are obtained with a neutral density filter 
 (attenuation of the intensity by factors of 80 in H-band and 90 in Ks-band)
and of a full aperture stop (instead of a 10\% undersized unobstructed circular stop).
The procedure for photometric calibration of such data is fully described in
\citet{Boccaletti08}. A detailed estimation of photometric
uncertainties accounting for the differences between out-of-mask and on-mask data leads to error bars of  0.16\,mag in both the H and Ks filters. The photometric error budget is actually driven by the flux extraction method of a point source (0.13\,mag) rather than the scaling factors.  The sources of errors being identical to those in \citet{Boccaletti08}, we will adopt the same photometric uncertainties. 


%
\subsection{Point source detection limit}
\label{sec:pointsource}

Finally, we measured the point-source detection limits on the images of Fig.\,\ref{fig:lyotraw} and Fig.\,\ref{fig:4qraw}. The 5-sigma contrast is calculated as the standard deviation of all pixels located at a given radius and that for each radius. 
Once converted to absolute magnitude, the result displayed in Fig.\,\ref{fig:detectionlimit} shows the detection limit for the H and Ks bands along the disk. Note that the 5-sigma detection level is taken here for convenience but does not rigorously correspond to a confidence level as usually defined for a Gaussian distribution. A detailed study of confidence level in the presence of diffraction residuals is given in \citet{Marois08}.

The FQPM image (Fig.\,\ref{fig:4qraw}) has a significantly worse detection limit in
the background dominated region ($>$2") owing to the 10\% transmission. 
However, the complementarity is clearly demonstrated in the limit of detection. 
The expected absolute magnitudes of giant planets (according to the DUSTY model of \citealt{Chabrier00}) 
for 5, 8, 10 and 20\,M$_{\mathrm{J}}$ were overplotted in Fig.\,\ref{fig:detectionlimit}. 

At 0.4", the 4QPM has the ability to pick out an object fainter than 8\,M$_{\mathrm{J}}$ while this level of sensitivity is achieved at 0.5" in the Lyot image. The candidate planet reported by \citet{Lagrange08} in Nov. 2003 data should have been detected at least in the 
K band image if located at the same position. However, our observations and those of  \citet{Lagrange08} are separated by nearly 1 year and hence orbital motion is to be expected. 

The projected distance of the planet candidate was 8\,AU ($411\pm 8$\,mas) on Nov. 2003. 
We therefore re-analyze our data to look around this position but found no firm evidence of a companion. However, a suspicious point-like pattern is visible in the Lyot image at $511\pm 18$\,mas i.e. $107 \pm 18$\,mas away from the discovery position (Fig. \ref{fig:circles}). It differs significantly from the \citet{Lagrange08} observations. This pattern intensity is compatible with the DUSTY model ($m_H \approx 13.5-14$). But no counterpart is detectable in the  4QPM image ($m_K(limit) \approx 13.4$) while the DUSTY model does not predict an equivalent flux in the H and Ks bands. Given the proper motion of \bp\ (about 82\,mas/yr to the north) a background star would appear to move closer to the star between 2003 \citep{Lagrange08} and 2004 (this paper). The source, if real in both epochs of data, moves farther away from the star between 2003 and 2004. Despite a lack of constraints on the orbital parameters,  a projected orbital motion of 107\,mas can be consistent with an actual separation larger than 8\,AU. However, the position angle of the suspicious point-like pattern differs by 4.5$^{\circ}$ with respect to the \citet{Lagrange08} observations which appears inconsistent with an orbit aligned with the disk. We therefore favor a false positive detection as an interpretation of the presence of this pattern.

Two hypotheses can be drawn from our 4QPM detection limit at Ks: 

1/ The physical separation in 2004 is less than 8\,AU and then the planet candidate is at an angular separation less than 0.4" which may account for the non detection in our data. In that case the motion is greater than 50\,mas. 

2/ The physical separation in 2004 is greater or equal to 8\,AU and the planet candidate is fainter than the model prediction by almost 1\,mag at Ks ($M_K(limit)\approx 12$). Parameters of the system (for instance age) must be reviewed and other models could be considered.
\\

It is difficult with the available material to identify the most plausible situation and we will have to wait for a definitive confirmation of the companionship and then carry out additional observations in H and Ks with a more effective rejection of the speckles. Orbital motion, if observable, will be valuable to derive the actual mass of the candidate planet and hence will provide a calibration of evolutionary models.

\begin{figure*}
  \centerline{\includegraphics[width=9cm]{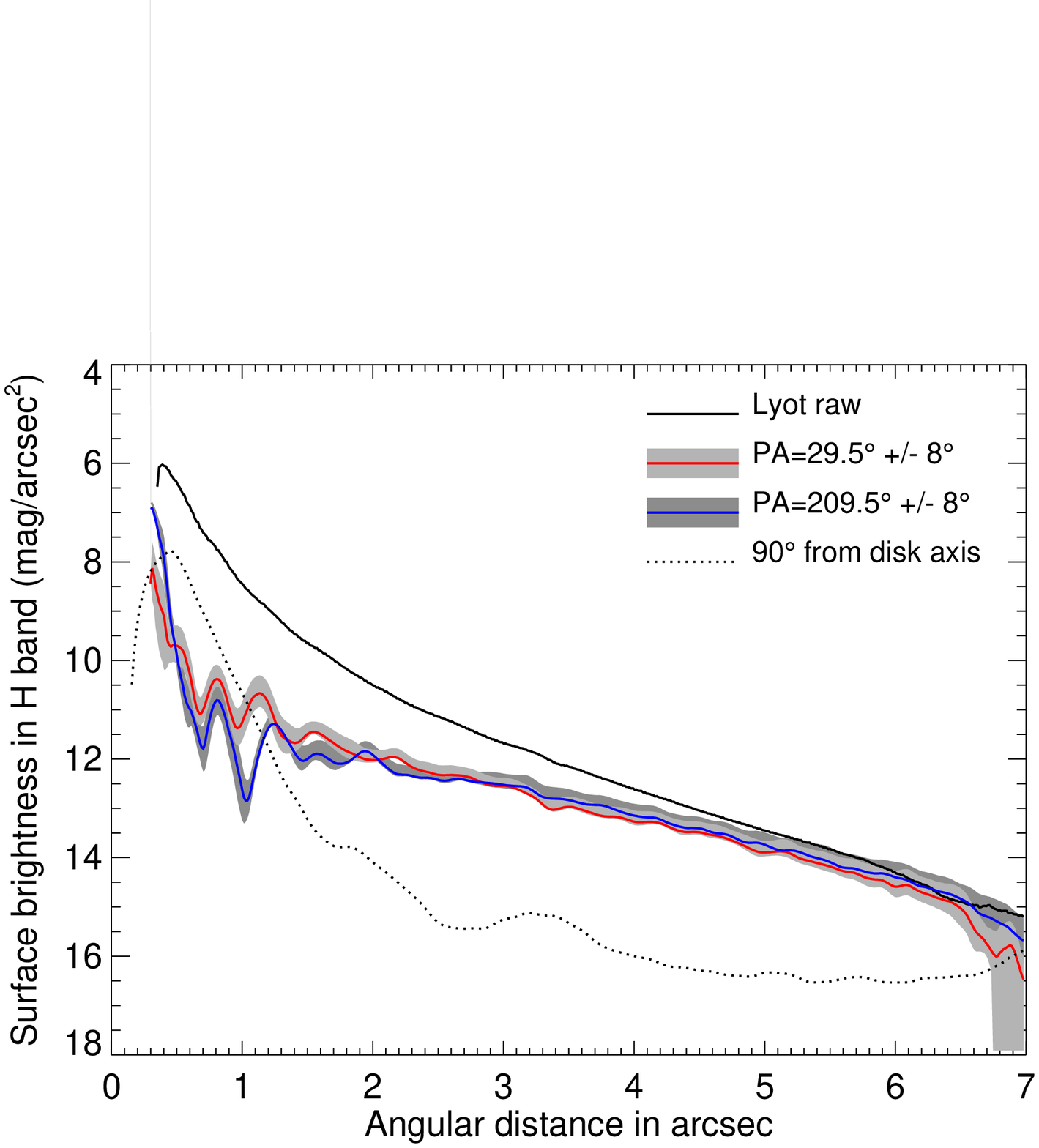}
    \includegraphics[width=9cm]{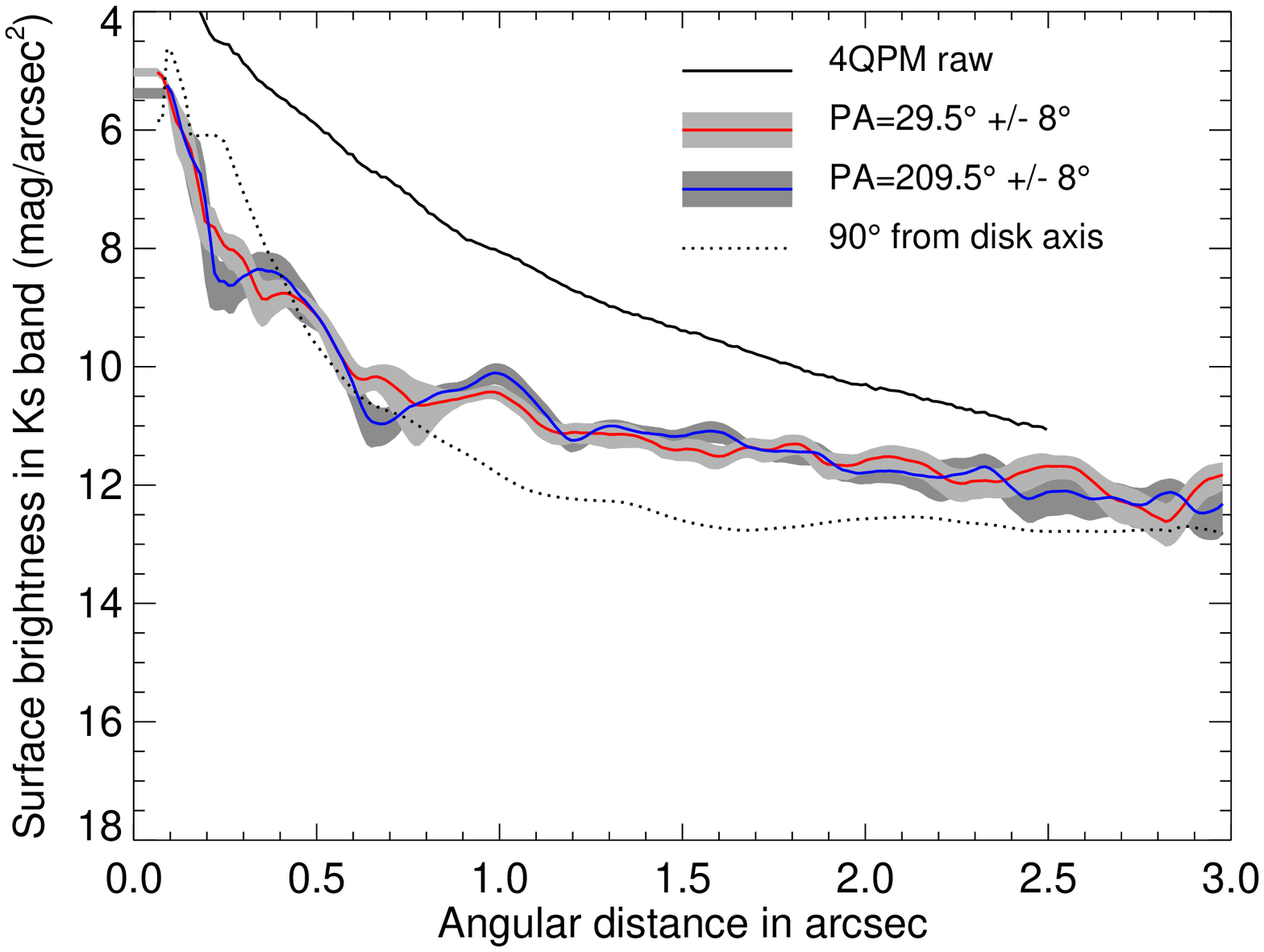}}
  \caption[]{Surface brightness of the disk (in mag/arcsec$^2$) versus
    the radial separation for the Lyot (left) and for the FQPM
    (right). The integration times are respectively 800\,s in the H band
    for the Lyot and 700\,s in the Ks band with a transmission of 10\%
    for the FQPM.   
    The solid lines show the radial profile of the raw image (before subtraction of a reference).
    The dashed and dash-dotted lines stand for the NE and SW sides, averaged over a 16$^\circ$ angular sector.
    The dotted lines give the noise level as measured at 90$^\circ$ from the disk axis.    
    The intensity ranges (1-sigma) in these sectors are  displayed as grey shades.
    }
  \label{fig:sb}
\end{figure*}
\begin{figure*}
  \centerline{\includegraphics[width=9cm]{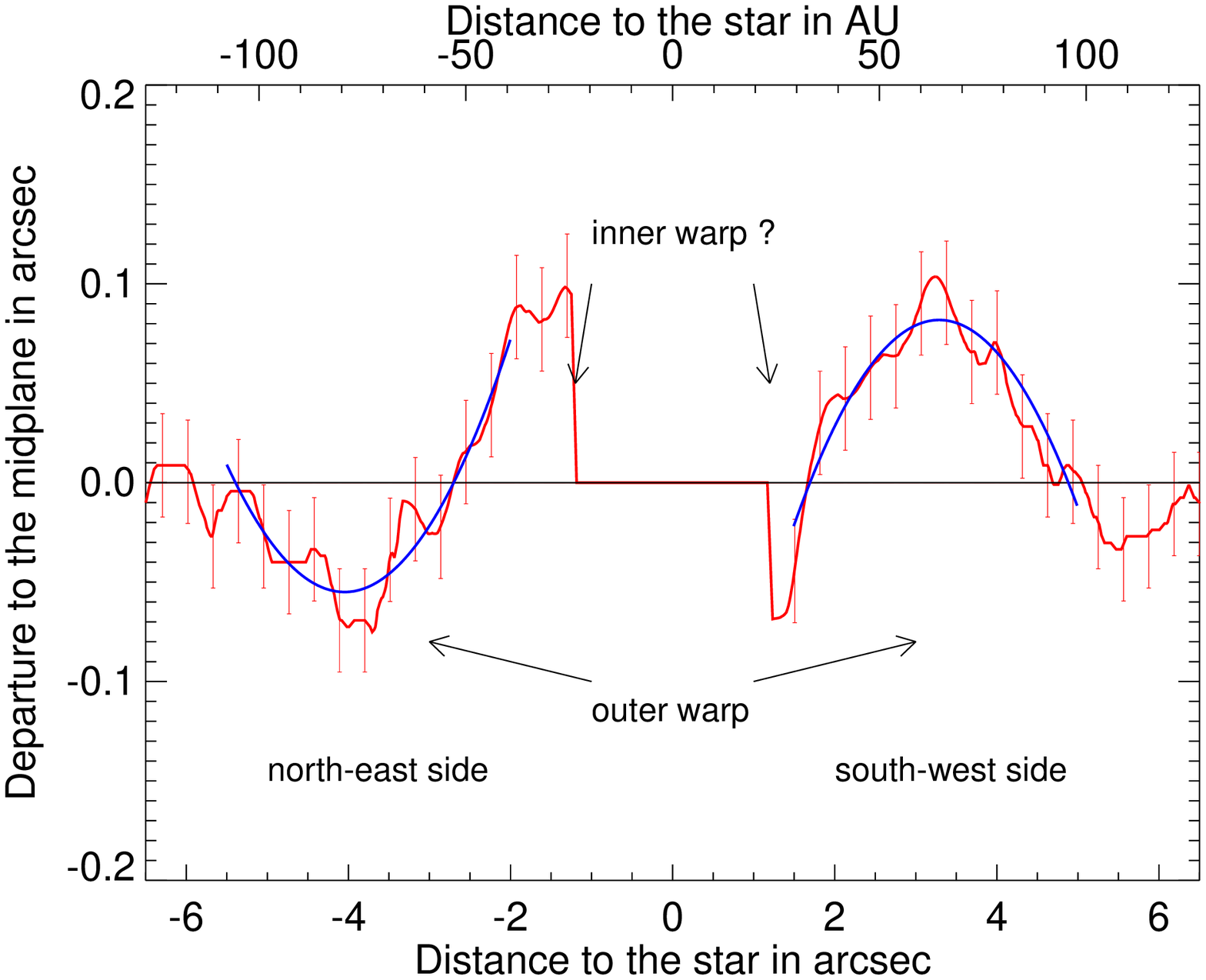}
    \includegraphics[width=9cm]{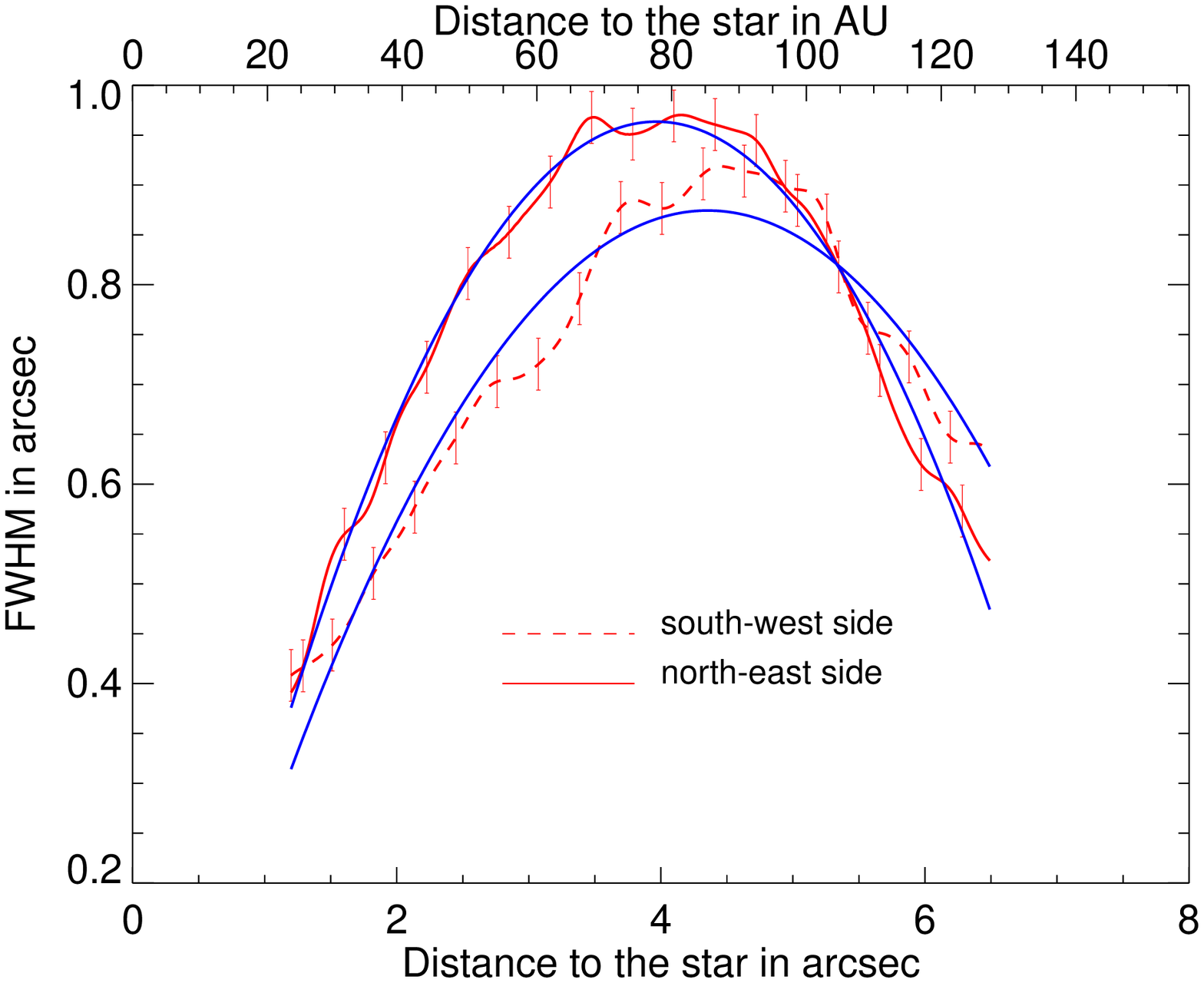}}
  \caption[]{Left: Departure of the disk spine with respect to the
    midplane as measured on the smoothed Lyot image. 
      Right: Thickness of
    the disk along the midplane. The FWHM of the two sides are
    overplotted to show the asymmetry. For these 2 plots, parabolic fits were used to derive positions of the asymmetries. Error bars are obtained from the same data with several smoothing factors relevant to the measurement of such spatial scales. } 
\label{fig:warp}
\end{figure*}

\section{Characterization of the dusty disk}\label{sec:analysis}
\subsection{Morphology of the disk}\label{sec:morph}
The \bp\ disk is well detected in scattered light in the Lyot image
(Fig.\,\ref{fig:horizontal}, top panel) up to 7.5" on the south-west (SW) side
and 7" on the north-east (NE) side (actually limited by the detector field of view). 
At such distances, the surface brightness is lower than 15/,mag/arcsec$^2$ (see section \ref{sec:sb}). 
To avoid confusion from diffraction
residuals either from the telescope or the coronagraph, most of the
radial features visible in Fig.\,\ref{fig:lyotraw} have been masked by
selecting a sector of about $30^\circ$ in the Lyot image about the
position angle (PA) of the disk (PA $=29.5^\circ\pm0.5^\circ$, measured from the north and counter-clockwise, see
section \ref{sec:mid}). The brightness asymmetry observed in the
visible by \citet{kalas95} is also seen in the near-IR (The NE side is
brighter than the SW side). The warp is also seen when comparing the
contour lines to the position of the midplane (dashed line in
Fig.\,\ref{fig:horizontal}). The largest deviation from the midplane is observed in
between 3" and 4" corresponding to a physical distance of 59-78\,AU, in
agreement with the measurement of HST/STIS \citep{heap00}.  The
innermost regions do not look aligned on the midplane but these
regions are highly dominated by residual diffraction patterns, as
explained in the next section.  Indeed, unlike HST/STIS images, the
image of the disk is not smooth but is perturbed by the radial
diffraction pattern from the primary mirror. This diffraction residual
produces ringed structures that we do not consider as real.

The FQPM image is displayed in the middle panel of
Fig.\,\ref{fig:horizontal} (a sector of 45$^\circ$ about the midplane was selected to
avoid confusion with instrumental radial features). To better
distinguish the large scale structures from the background noise we
applied a Gaussian filtering to the image shown in
Fig.\,\ref{fig:4qraw} (right). As explained in section \ref{sec:obs}, the
signal to noise is worse than in the Lyot image, despite similar
integration times, since the transmission is 10 times lower. The disk
is detected down to a level of 12.5~mag/arcsec$^2$ at a radius of
about 2".  Although the central field is not obscured as in the Lyot
image, a discontinuity at 0.7" on both sides suggests that the
patterns located below 0.7" are actually diffraction residuals. This
is confirmed by the measurement of the surface brightness in section
\ref{sec:sb}. The warp is obviously not detected due to a lack of
sensitivity at distances larger than 2".  However, a very pronounced
pattern resembling a knot at about 1" is clearly visible on the SW
side. It was not seen in the Lyot image owing to the presence of a
diffraction spike and its origin remains unexplained. 
This feature surprisingly resembles the clump observed by \citet{telesco05} in the mid-IR although not located at the same position (52\,AU in the SW side), while we are probing closer separations (1" is equivalent to about 20\,AU).  Other clumps (beyond 50\,AU) reported by \citet{wahhaj03} and \citet{weinberger03} in the mid-IR are interpreted as the projection of rings since they are symmetrical. Such rings are also observed at much larger separations ($>$500\,AU) in the visible by \citet{Kalas00}. However, in our near-IR data, the clump has no symmetrical couterpart in the NE side and the origin must be found elsewhere than in the presence of rings. The refinement of the planet candidate orbit \citep{Lagrange08} will certainly allows to perform more accurate dynamical simulations and possibly be helpful for the interpretation of such patterns.

A smooth contour of the Lyot image is overplotted on the FQPM
image. The overlap is only partial because the usable field is
different for each coronagraph (positions of the spikes and coronagraph signatures are different).

\begin{figure*}
  \centerline{\includegraphics[width=9cm]{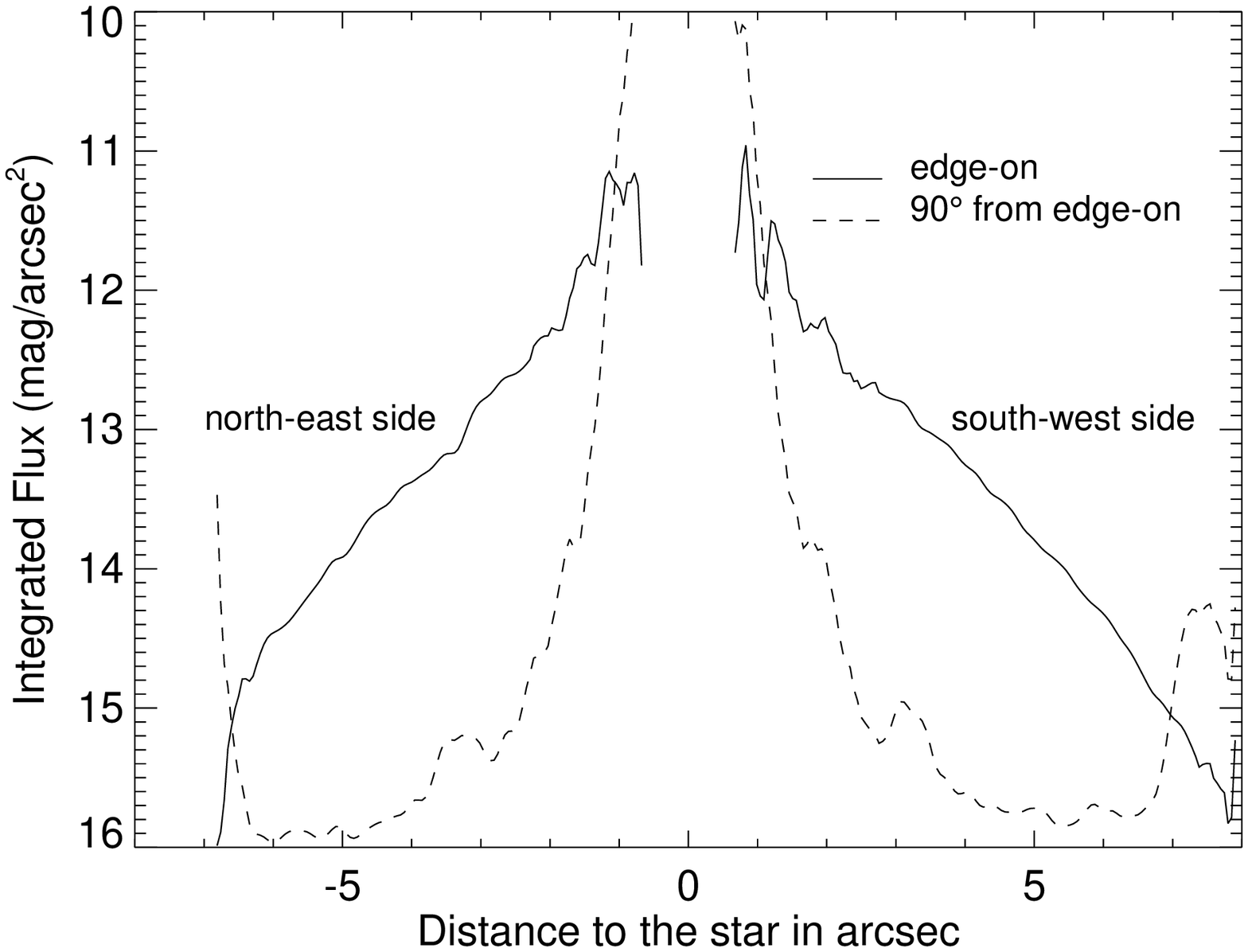}
    \includegraphics[width=9cm]{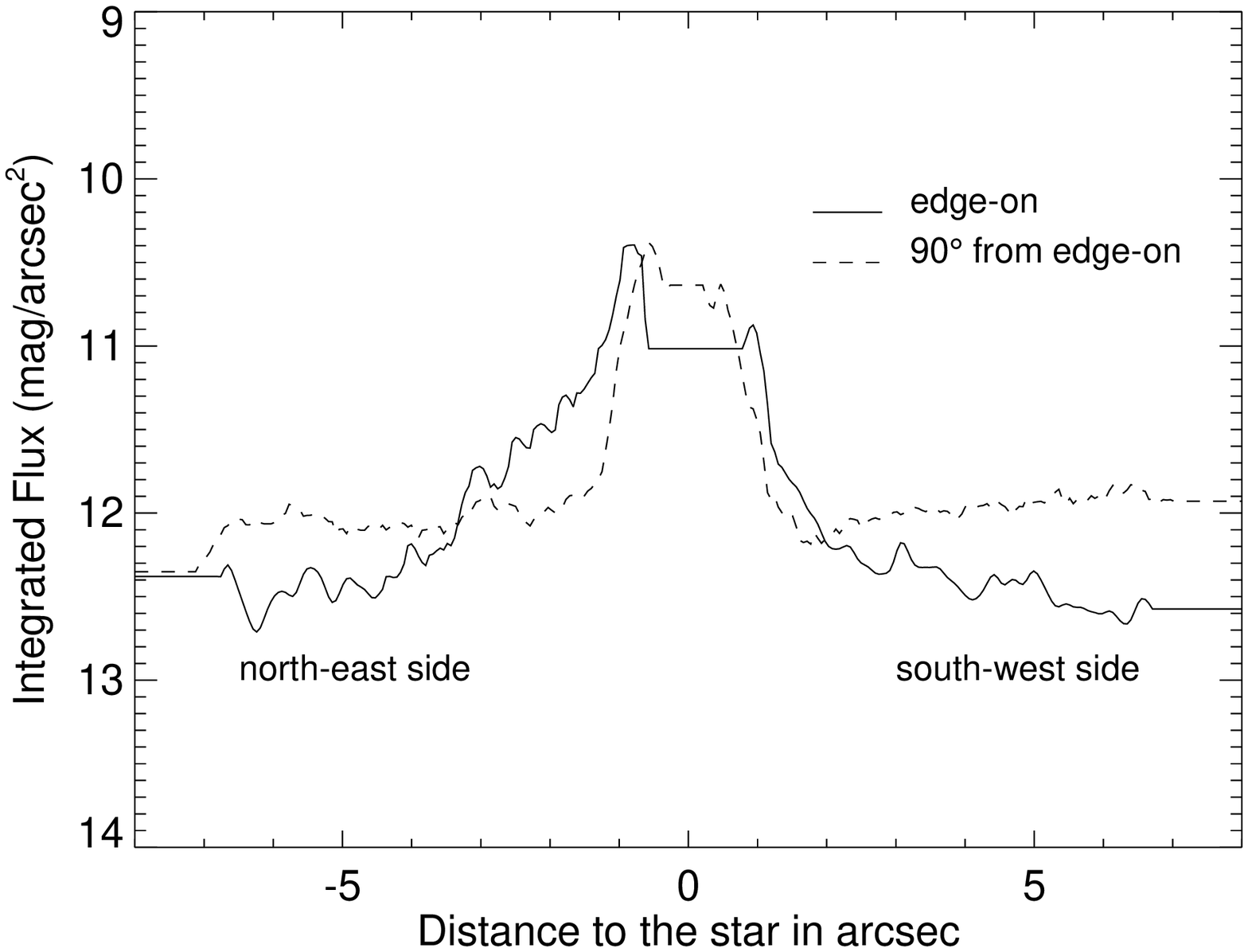}}
  \caption[]{Vertically integrated flux as a function of the angular
    separation to the star for the Lyot (left) and the FQPM
    (right). Dotted lines correspond to measurements at
    90$^\circ$ from the disk axis. 
     }
  \label{fig:fluxvertical}
\end{figure*}

\subsection{Surface brightness}
\label{sec:sb}
The surface brightness (SB) of the disk was measured along the NE (PA
$=29.5^\circ$) and SW (PA $=209.5^\circ$) sides and averaged over a
$\pm 8^\circ$ sector (avoiding too many diffraction residuals). The SB
radial profiles derived from the Lyot and the FQPM images are shown in
Fig.\,\ref{fig:sb} (left and right panels, respectively). The
$1\,\sigma$ intensity variation versus radius is displayed with grey
shades. The surface brightness of the disk is compared to the
brightness measured in a similar sector but perpendicular to the disk axis (a region free of circumstellar material) to estimate an
equivalent noise level. The modulus of this level is plotted in Fig.\,\ref{fig:sb} as a dotted line.
In the Lyot image, the region closer than $\sim$1.2" (23.2\,AU) is totally
dominated by diffraction residuals although the coronagraphic mask is
only 0.35" in radius.  In the FQPM image, the field is not obscured in
the center but the diffraction residuals are clearly stronger than the
disk brightness for separations closer than $\sim$0.7" (13.5\,AU).  At
large distances ($>$2.2"), the FQPM image is rapidly dominated by the
background noise, as expected.

On the NE side, the two profiles are similar within the intensity range (grey shades), while on the SW side a small difference appears near 1.5". However, it is difficult to interpret this difference as a color effect
because the coronagraphs have different signatures and the overlap of the detected regions is only partial. 
Also, it is not sufficient to constrain the grain size distribution as we did for HD\,141569 in \citet{boccaletti03}.

In both the Lyot and FQPM cases, the diffraction residuals in the
center can probably be improved with a more careful matching of the
star and reference images during the observation (better centering and
better correction of the low order aberrations are needed). The
advantage of the FQPM over the Lyot coronagraph, although less
sensitive in the background dominated region, is clearly seen
here and allows us to detect the disk at a separation which is almost
half the size than on the Lyot image. This compares to the best
distance reported in the literature and achieved with the HST/STIS
instrument, where the disk is detected down to 0.75"
\citep{heap00}. However, the exact signal-to-noise ratio at these
distances cannot be estimated on the STIS images because
\citet{heap00} do not compare the average brightness of the disk with
the noise level.

\subsection{Spine and thickness}
\label{sec:mid}
The position angle of disk midplane
is measured on the Lyot image by a line passing through the star and
going from the edge of the NE side of the disk to that of the SW
side. The disk PA is found to be $29.5^\circ\pm0.5^\circ$, consistent
with previous measurements \citep[between $30.1^\circ$ and
$31.4^\circ$ according to ][for separations larger than $\sim
3"$]{kalas95}.

In the inner disk regions, the known warp makes the disk not perfectly
aligned with the midplane, but instead it appears twisted as shown for
instance by \citet{mouillet97b} and \citet{heap00}. Along cuts
orthogonal to the midplane, the distance of the peak brightness
position to the midplane defines the disk spine which is displayed in
the left panel of Fig.\,\ref{fig:warp} for our Lyot image and for a
separation larger than $1.2"$. This figure was obtained in two
steps. First, the image of the disk was convolved with a Gaussian
function to account for the sole large-scale structures. Then, we
measured the altitude of the peak brightness with respect to the
midplane as a function of the distance from the star using two
different methods: one consists of directly measuring the distance of
the maximum intensity to the disk midplane, while in the second case,
the spine position is obtained from a Gaussian fitting to the vertical
profile. Both methods yield very similar results.

According to Fig.\,\ref{fig:warp} (left panel), the position of the
warp in the SW and NE sides appears to be not symmetrical with respect
to the star. The warp extends to 3" on both sides but peaks at about 3.3" (65\,AU) on the SW side and
4.0" (78\,AU) on the NE side as measured with parabolic fits
(overplotted on Fig.\,\ref{fig:warp}). This shift cannot be attributed to a
mis-alignment of the star behind the mask, since the centering
accuracy is better than 0.1".  The difference in the position of the
warp with respect to the star is not reported in previous
observations.  It is reminiscent of the disk offset discovered in
Fomalhaut by \citet{Kalas05} and to a lesser extent in the disk around
\object{HD\,141569} \citep{boccaletti03, mouillet01}. The amplitude of
this offset is physically almost identical in the cases of \object{Fomalhaut} and
\bp\ (about 15\,AU), although the \bp\ offset might be larger due to
projection effects. It has been suggested that a planetary companion
on an eccentric orbit ($e=0.02$) could create such asymmetries
\citep[e.g. \object{HR\,4796,}][]{Wyatt99}. However, it is noted that the mass of the pertubating planet is not very well constrained since several combinations of mass and semi-major axis are able to reproduce the observations. 
The recent detection of a planet around Fomalhaut \citep{Kalas08} confirms the relationship between the perturber and the eccentricity of the ring. The most stringent study in the case of \bp\ is presented by \citet{Freistetter07}. To reproduce the 
planetesimal belt positions inferred from high-resolution mid-IR spectroscopy, a system with three planets is necessary, one of them being significantly more massive ($2 ^{+3}_{-0.5}\,M_J$) and located at 12\,AU. 
A revised value of the \bp\ age \citep[12\,Myr,][]{Zuckerman01} together with the reasoning of \citet{heap00} 
and an averaged position of 71 AU of the warp yield a mass of about 2\,$M_J$, similar to that of \citet{Freistetter07}. Our detection limit presented in Fig. \ref{fig:detectionlimit} is able to place an upper mass limit of 5-6\,M$_J$ at 10\,AU for an hypothetic planet of 10\,Myr. This is identical to the upper limit of \citet{Freistetter07} and therefore not relevant to improve the mass constraint  while the planet candidate from \citet{Lagrange08} is actually more massive (8\,$M_J$).

The altitude of the maximum deviation with respect to the
midplane reaches 0.10" (1.96 AU) in the SW and 0.07" (1.37 AU) in the
NE extension. \citet{heap00} also observed a higher altitude in the SW
side although they did not discuss this point. Another deviation
from the midplane is also visible at closer separations 
 and is located at 1.1" and 1.5" respectively on the SW and NE sides. This
inverse warp is likely an artifact since we do not detect it on the FQPM
image which is more accurate than the Lyot image in this
region. We suspect that the confusion near the center between
circumstellar material and diffraction residuals from the on-axis star
is responsible for such patterns which are enhanced by the use of a
Lyot coronagraph. It might be the same kind of artifact that lead
\citet{heap00} and \citet{golimowski06} to conclude that the two
extensions in the inner disk do not intersect at the star. 

 \begin{table}[t]
\caption{Registration of asymmetries in the warp of \bp\ .}
\begin{center}
\begin{tabular}{lcc} \hline\hline
							&	North-East	&	South-West \\ \hline
position along midplane (AU)		&	78			&	65 \\
elevation (AU)					&	1.4			&	2.0\\
maximum thickness position (AU)	&	78			&	87\\ 
maximum thickness (AU)			&	18			&	18\\ \hline
\end{tabular}
\end{center}
\label{tab:asym}
\end{table}%

The Gaussian fit also yields the full width at half maximum (FWHM)
of the disk along the midplane.  The maximum thickness occurs at about
4.5" and 4.0" from the center respectively for the SW and NE sides
(Fig. \ref{fig:warp}, right). This asymmetry is opposite to that of
the warp position. In addition, the SW side is marginally thinner than
the NE side (at 1-sigma) a characteristic also reported at mid IR wavelengths
\citep{pantin97}. 
The average thickness is about 0.9" (18 AU), in perfect agreement with \citet{heap00} and again with \citet{pantin97}. There is a consensus for this parameter at visible, near IR and mid IR wavelengths. 
In \citet{golimowski06}, the SW side appears thinner although error bars are missing to confirm this point. However, they claim that
actual thicknesses are 30$\%$ less after PSF deconvolution. However,
the narrow HST/ACS PSF (FWHM=0.05") cannot increase the apparent size
of a pattern which is more than 15 times larger. With a simple
numerical simulation, we measured that a disk of which the FWHM is 15
times larger than the PSF size undergoes an increase of only 0.5\% in
size once convolved with the PSF. Clearly, the impact of
PSF-deconvolution in coronagraphic imaging is a very critical issue
and will be discussed in section \ref{sec:deconv}.

A summary of the asymmetries we measured inside the disk is reported in Table \ref{tab:asym}. 

\begin{figure}
  \centerline{\includegraphics[width=9cm]{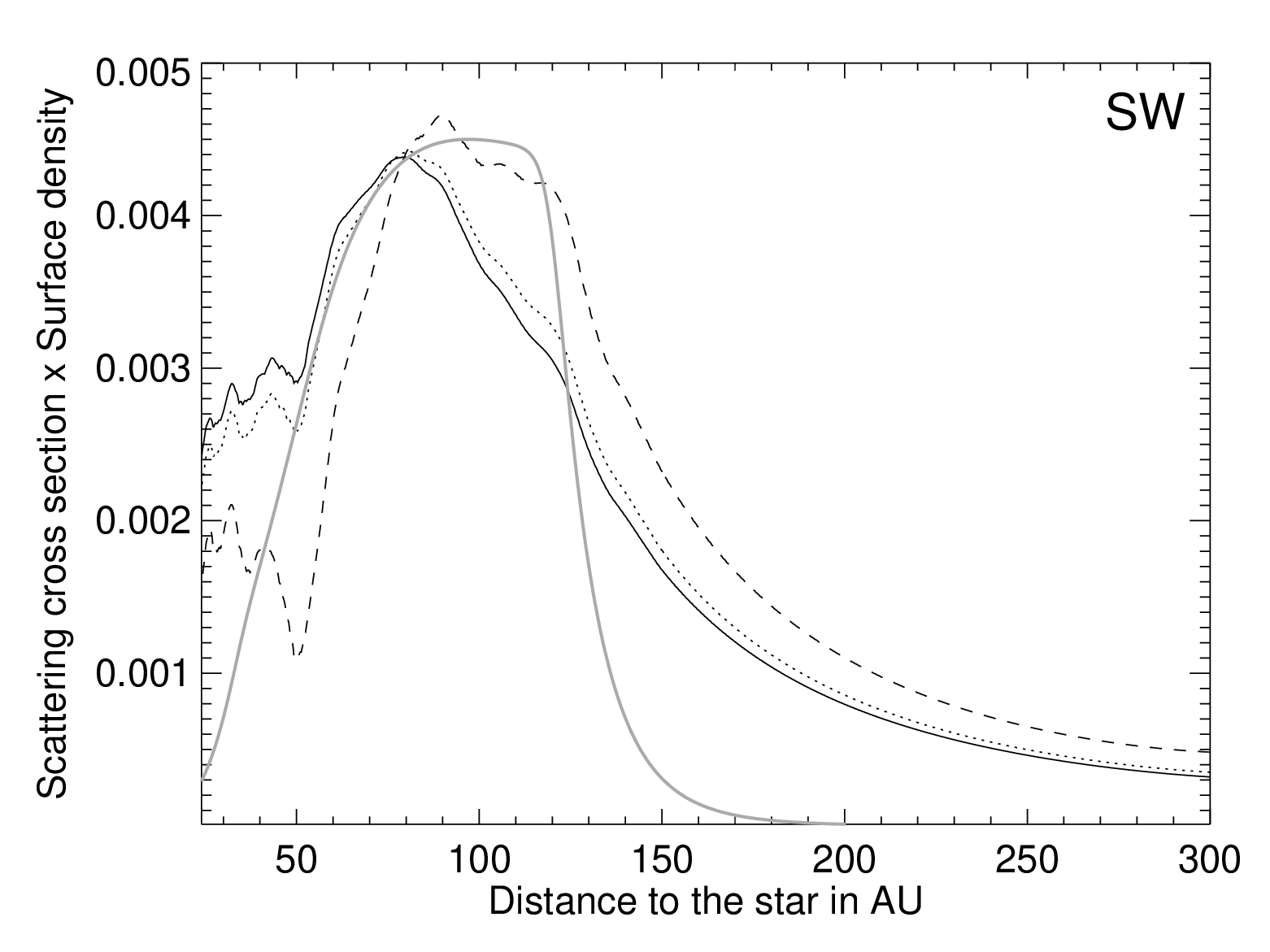}}
  \centerline{\includegraphics[width=9cm]{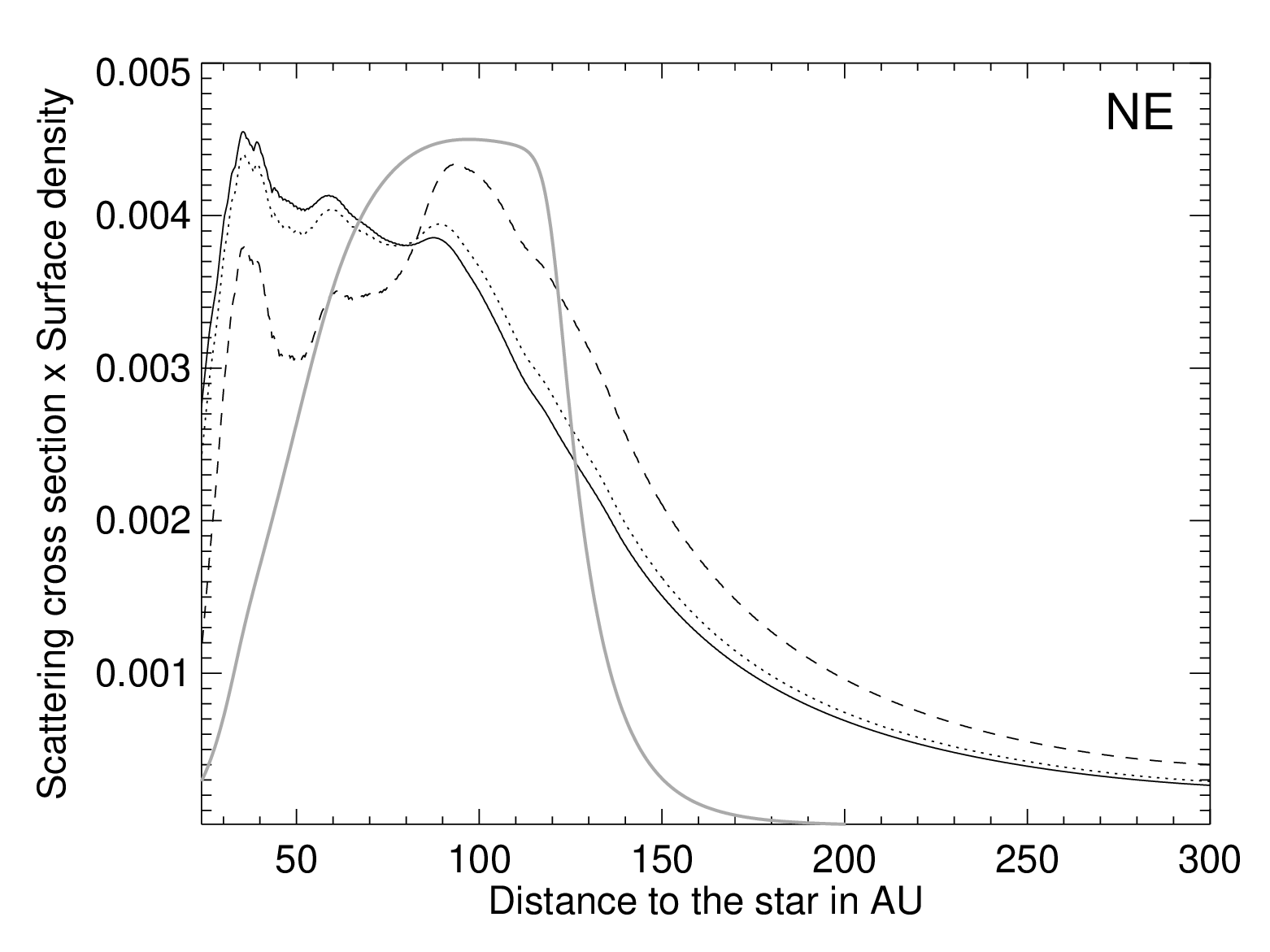}}
  \caption[]{$\sigma_{\mathrm{sca}}^H\Sigma(r)$ {\it dust} disk
    profiles inferred from the inversion of the Lyot vertically
    integrated SW and NE profiles diplayed in
    Fig.\,\ref{fig:fluxvertical}, and assuming 3 different scattering
    asymmetry factors: $|g|=0$ (solid line), $|g|=0.25$ (dotted line)
    and $|g|=0.5$ (dashed line). The gray solid line shows the {\it
      planetesimal} profile (normalized to 0.0045 at its peak
    position) inferred by \citet{augereau01} for comparison.}
  \label{fig:inversion}
\end{figure}

\begin{figure*}
  \centerline{\includegraphics[trim= 5mm 5mm 5mm 5mm ,clip,width=15cm]{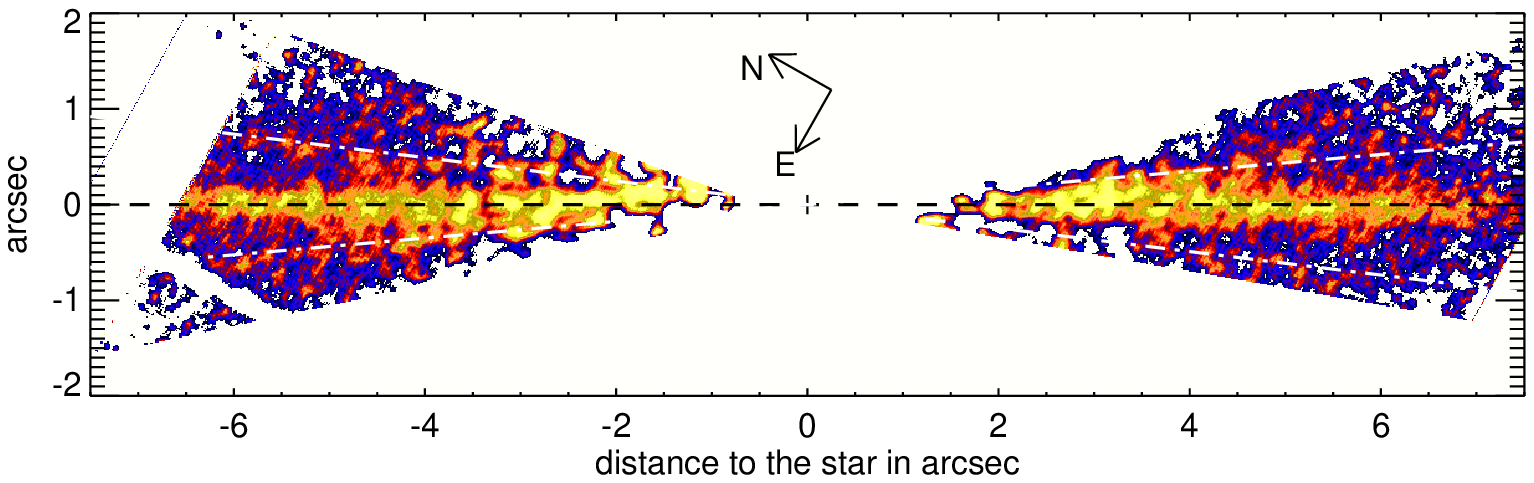}}
  \centerline{\includegraphics[trim= 5mm 5mm 5mm 5mm ,clip,width=15cm]{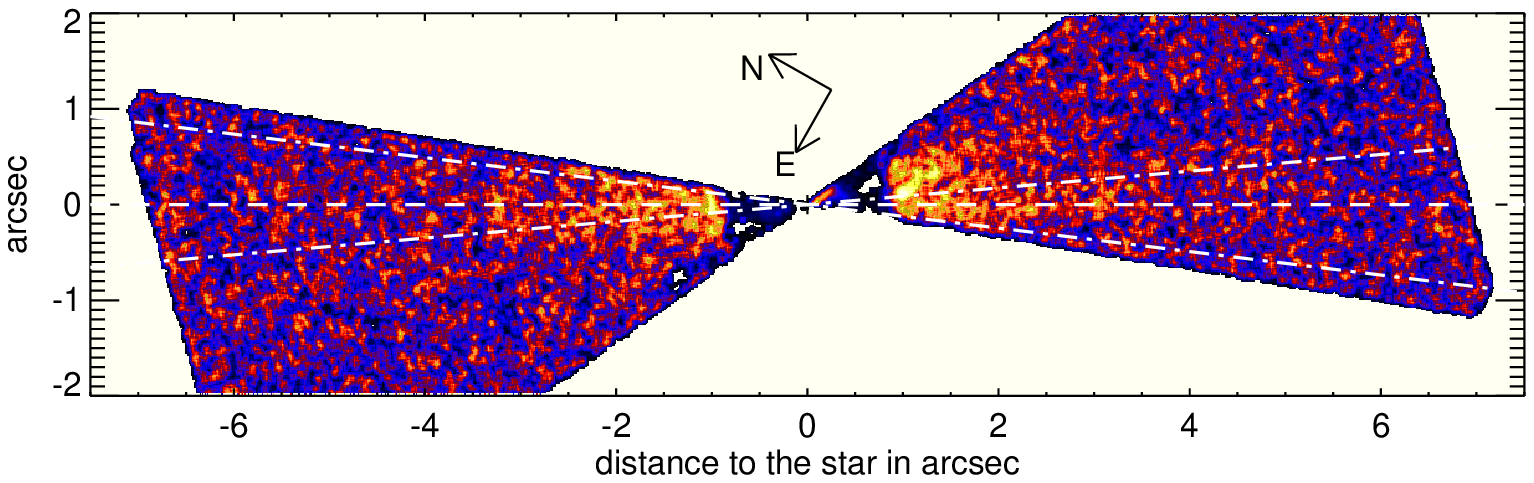}}
  \caption[]{Figures identical to Fig.\,\ref{fig:horizontal} (top and middle) but after deconvolution and
    division by the original, non-deconvolved image. Two dashed lines
    at $5^\circ$ and $-7^\circ$ from the main disk trace the
    observed radial structures in the Lyot image. 
   }
  \label{fig:bpicdeconv}
\end{figure*}

\subsection{Vertically integrated flux and normal optical thickness}
\label{sec:vertical}
We finally estimate the vertically integrated flux which is measured
across the disk thickness ($z$-axis) and shown as a function of the
radial distance to the star in Fig.\,\ref{fig:fluxvertical}. This flux
is obtained by integrating the disk surface brightness over a
  $\Delta z=1$'' region centered on the disk midplane, and normalized
  to the integrating area.
Residual patterns about the center are avoided with an appropriate
mask which is adapted to each coronagraph.  The vertically integrated
flux is almost symmetrical for the NE and SW sides in the Lyot image. This is not the case with the FQPM. 
Again, the advantage of the Lyot is clearly seen at large separations ($>$2"), but the FQPM image is still useful to rule
out diffraction artifacts at separations closer than 2" and down to
0.7".  The same measurement performed at 90$^\circ$ from edge-on
indicates the regions where we can be confident in the value of the
vertically integrated flux.  
As we took the modulus of the residuals (to avoid negative values), a numerical effect makes the 
 equivalent noise level sometimes larger than the flux measured inside the disk. This is particularly pronounced with the FQPM.

The disk being seen edge-on, the observed surface brightness profile
at any given projected distance $r_{\mathrm{proj}}$ is the sum of the
brightness contributions, modulated by the scattering phase function,
of all the grains physically located at distances $r \geq
r_{\mathrm{proj}}$ along the considered line of sight. Deriving the
shape of the surface density profile from an edge-on disk brightness
profile requires the use of a deprojection method, such the one
employed by \citet{augereau06} for \object{AU\,Mic} 
or the numerical one described in \citet{pantin97} for \bp\.
 In this approach, the product of the dust surface density profile $\Sigma(r)$ and the
mean scattering cross section of the grains $\sigma_{\mathrm{sca}}$ is
reconstructed from the outer disk edge down to the closest distance.
For the inversion process to work properly, smooth profiles that
extend to large distances where the disk brightness vanishes are
required. It is therefore not well adapted to the FQPM profiles, but
more suited to the Lyot surface brightness profiles which cover a much
larger range of distances. Before performing the inversion, the Lyot
profiles have nevertheless been extrapolated 
beyond $140$\,AU with a $r^{-4.5}$  radial profile and a Gaussian vertical profile (FWHM$(r)$ taken from 
\citealt{golimowski06}), and then smoothed over $10$ pixels.  

Examples of deprojected normal disk profiles using the inversion
technique of \citet{augereau06} are displayed in
Fig.\,\ref{fig:inversion} for the SW and NE sides. The figure shows
the product of the mean scattering cross section of the grains in the
H-band $\sigma_{\mathrm{sca}}^H$ by the dust disk surface density
$\Sigma(r)$ for different scattering asymmetry factors $|g|$ between
$0$ (isotropic scattering) and $0.5$.  The
$\sigma_{\mathrm{sca}}^H\Sigma(r)$ profiles are equivalent to normal
dust disk optical thickness profiles multiplied by the mean albedo of
the grains in the H-band. The \bp\ disk appears $2$ to $4$ times less
dense at the peak position than the $2.5$ times more distant annulus
about the F5/F6\,V star \object{HD\,181327} that peaks around
87\,AU \citep{schneider06}.

Beyond $\sim150\,$AU, the calculated profiles follow a $r^{-2.5}$
power law as expected from the extrapolated surface brightness
profiles at these distances. The SW $\sigma_{\mathrm{sca}}^H\Sigma(r)$
profiles show a peak, whose exact position varies with $|g|$, but
is situated in the range $80$--$90$\,AU. The global ring-shape of the SW
dust disk is consistent with the normalized (to allow comparison)
planetesimal surface density inferred by \citet{augereau01} (gray
thick line in Fig.\,\ref{fig:inversion}). The dust profiles in
Fig.\,\ref{fig:inversion} appear more extended than the theoretical
planetesimal belt because the radiation pressure force causes the
grains produced through collisions to populate the outer disk regions
as shown in \citet{augereau01}. The NE
$\sigma_{\mathrm{sca}}^H\Sigma(r)$ profiles below $100$\,AU do not
mirror the SW profiles suggesting a non-axisymmetrical structure.  The
NE profiles indeed appear flatter than the SW profiles, and the peak
at around $90$\,AU only becomes visible for very anisotropic
scatterers (large $|g|$ values). It is also noticeable that the observations
in the mid-infrared deduced surface density profiles \citep{pantin97} are 
consistent with $|g|$=0.5 profiles estimated here.

Although the exact shape of the disk
cannot be derived uniquely (because of uncertainties in the
observations and in the scattering properties in particular), both the
NE and SW $\sigma_{\mathrm{sca}}^H\Sigma(r)$ profiles point toward the
presence of dust material inside the main $80$--$100$\,AU planetesimal
belt, as previously suggested for example by mid-IR observations
\citep[][and ref. therein]{telesco05}, and by the present FQPM
images. Overall, the results tend to support the idea of a complex,
structured and asymmetric system inside the apparent warp location,
which led \citet{Freistetter07} to conclude the presence of several
planetary companions.

\section{On the warp}
The warp of \bp\ has been the subject of many investigations and
hypotheses especially regarding the presence of pertubing bodies
like planets \citep{mouillet97b, heap00, augereau01,
  Freistetter07}. \citet{heap00} and later \citet{golimowski06} have
interpreted the warp as the blend of two distinct disks at slightly
different position angles instead of a single distorted structure.
To emphasize the presumed secondary disk, \citet{golimowski06} used a
reduction process that we reproduce in Sec.\,\ref{sec:warp} and
critically discuss in Sec.\,\ref{sec:deconv}.

\subsection{The secondary disk}\label{sec:warp}

The coronagraphic images presented in \citet{golimowski06} were
obtained with the HST/ACS instrument and classically subtracted with
the image of a reference star. In this paper we adopted a similar
approach for the data reduction (Sec.\,\ref{sec:reduction}). This
processing is extremely efficient in the case of the HST owing to the
stability of the PSF compared to that of ground-based
telescopes. Then, the HST/ACS subtracted coronagraphic image was
deconvolved with a standard Lucy Richardson algorithm taking a
synthetic Tiny Tim \citep{Krist04} image as a PSF.

In classical imaging, the image results from the convolution of the
object by the PSF. Deconvolution assumes that the PSF is stationary (independent
of its position in the field). This is no longer the case in coronagraphy since the image variation is strongly
non-linear especially near the center \citep{malbet96}. The term PSF is no longer applicable. 
In a coronagraph, the convolution occurs in the pupil plane downstream of
the mask where the pupil is convolved by the Fourier transform of the
mask \citep{malbet96}. A PSF defined for each point in the image might
be necessary in this case for deconvolution.  \citet{golimowski06}
have worked around the problem by avoiding the inner region inside a
$1.5$" radius, but the effect of deconvolution on coronagraphic image
is not further analyzed.  At this stage, the warp is still visible as a
structure in the main disk. To improve the visibility of fine
structures, \citet{golimowski06} have divided the deconvolved
 with the non-deconvolved subtracted coronagraphic image. This
procedure reinforces the high spatial frequencies of the images (as in
unsharp masking) and reveals the warp as two separate disks. This
hypothetic secondary disk can be distinguished from the main disk
beyond $80\,$AU. Below that distance, the intrisic disk thickness
prevents any partition of the two disks.

However, this structure is difficult to interpret. It could be a
secondary disk fully separated from the main disk and centered on the
star, or could instead originate from the main disk as for instance
suggested by the inward extrapolation of NE secondary disk which
intercepts the main NE disk at around $30\,$AU
\citep{golimowski06}. The surface brightness profile of the secondary
disk as measured by \citet{golimowski06} is very reminiscent of the
brightness profile of the main disk beyond $125\,$AU. The latter
distance is supposed to represent the outer edge of the main
planetesimal belt from which the dust grains originate that populate
the outer disk regions due to radiation pressure forces
\citep[e.g.][and ref. therein]{augereau01}. Similarly, the secondary
disk could represent the dusty tail of an inclined population of
planetesimals located within $80\,$AU.
It is noteworthy that the extrapolation inward of the secondary disk surface
brightness measured by \citet{golimowski06}  would make it
brighter than the main disk below $70\,$AU, which would make the main
disk brightness profile much steeper than actually observed. The
rather flat surface brightness profile observed below $4$'' suggests
that the secondary disk brightness profile breaks at around
$70$--$80\,$AU and becomes flatter inside that distance. The reasoning
developed for the main disk may thus also be valid for the secondary
disk, which would then originate from parent bodies arranged in the
form of an inclined belt with an outer edge close to
$70$--$80\,$AU. It is remarkable that this distance matches the
maximum extension for the warp produced by the inclined planet in the
models of \citet{mouillet97a} and \citet{augereau01}.

Our reduced coronagraphic images were deconvolved as in
\citet{golimowski06} except that we were using an actual PSF image of
\bp\ (out of the coronagraph) and a maximum likelihood algorithm for
the deconvolution. As a result of the deconvolution process, we
marginally detect a radial structure with an inclination of $5^\circ$
with respect to the disk midplane and consistent with the secondary
disk, but also one almost symmetrical structure about the disk
midplane at $-7^\circ$ from the main disk (Fig.\,\ref{fig:bpicdeconv},
top panel). It is however difficult to measure the radial distance
where these structures originate, although they are consistent with
purely radial patterns 
intersecting the star unlike the
secondary disk (at least on the NE side) observed with the ACS
instrument. Here, we interpret the radial  patterns revealed by the
deconvolution process as PSF radial  patterns instead of real
patterns. These patterns are therefore different to those observed by \citet{golimowski06}. Our emphasized FQPM image is presented in the bottom panel
of Fig.\,\ref{fig:bpicdeconv} but does not show any features.

\begin{figure}[ht]
  \centerline{\includegraphics[width=9cm]{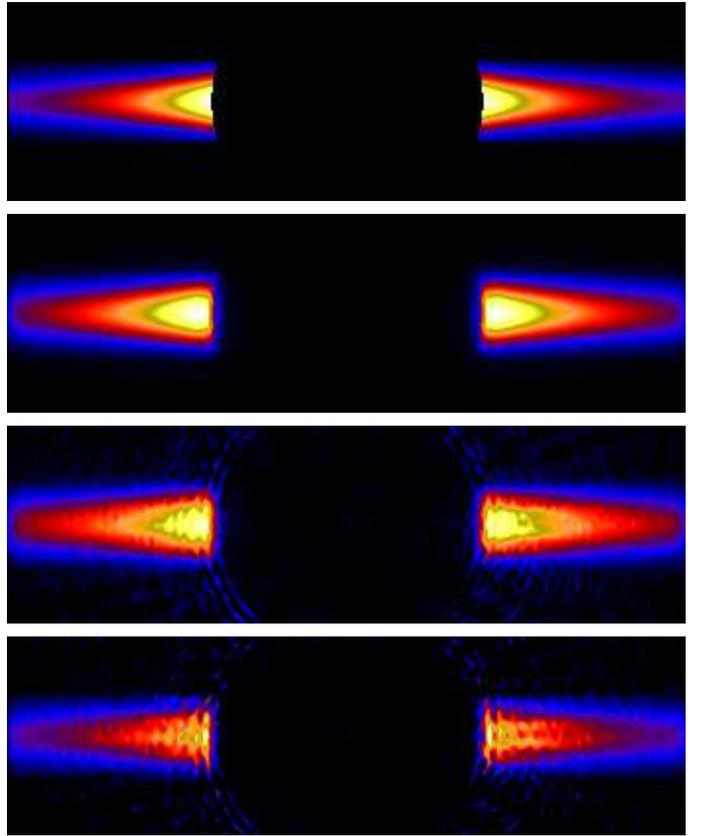}}
  \caption[]{Simulated images of (from top to bottom), the disk model, 
  the disk convolved with the PSF, the coronagraphic image
    subtracted with that of a reference star and the deconvolved
    image. 
   The FOV is $4.5" \times 1.3"$.}
  \label{fig:deconv_ima}
\end{figure}

\subsection{Deconvolution of coronagraphic
  images: simulation }\label{sec:deconv}

\begin{figure*}
  \centerline{\includegraphics[width=9cm]{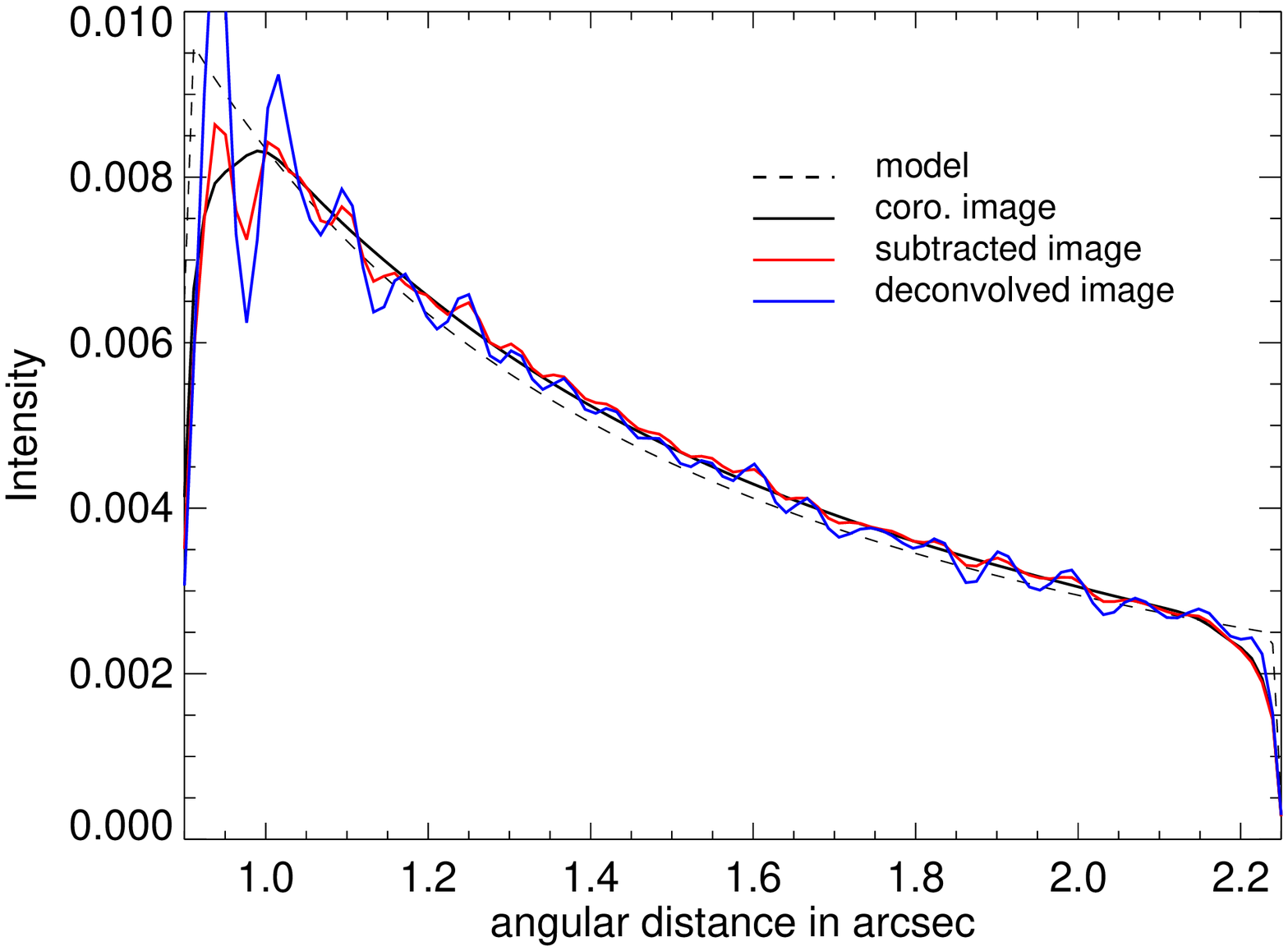}
    \includegraphics[width=9cm]{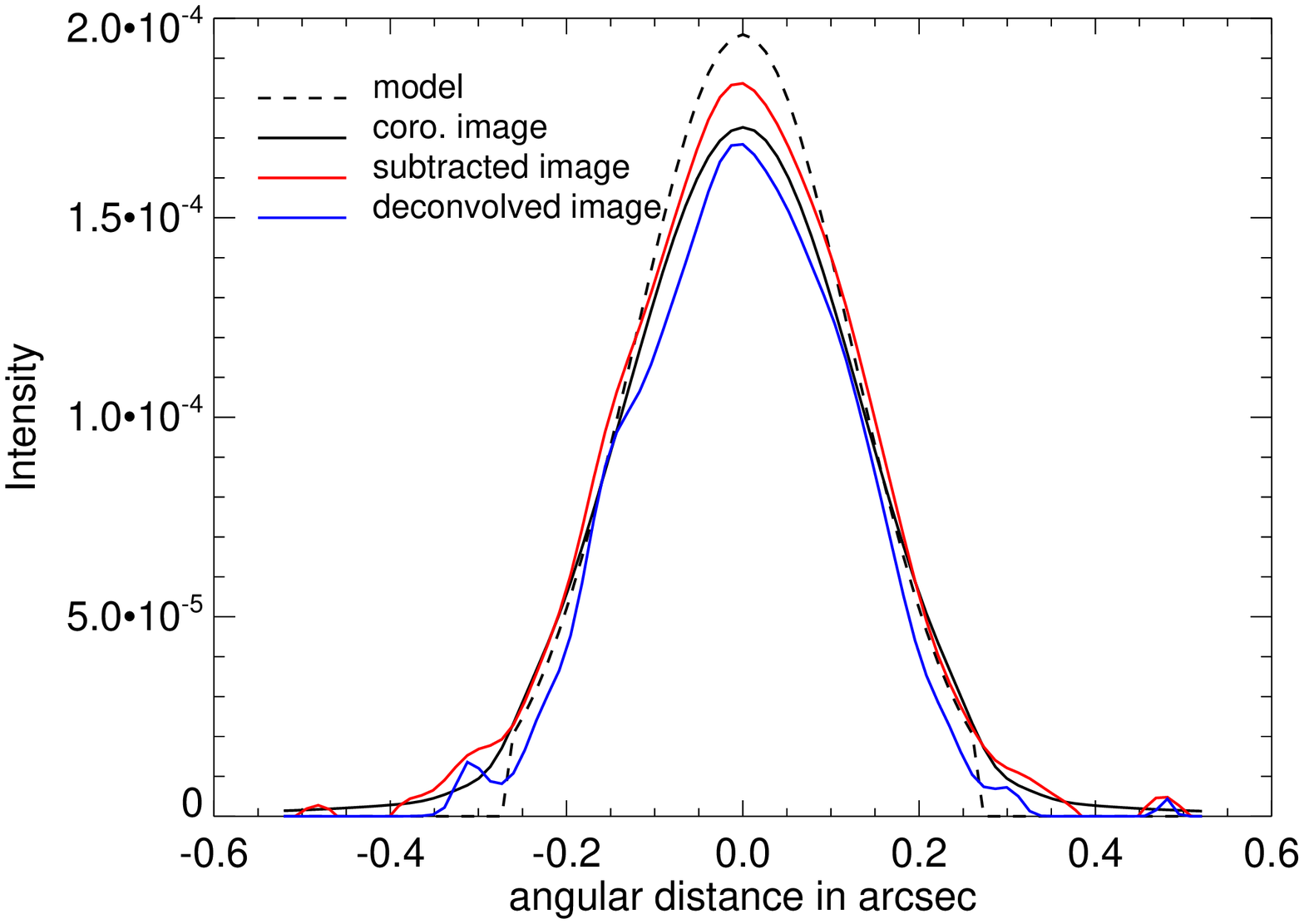}}
  \caption[]{Comparison of the radial intensity profile (left) and the
    vertical profile of the disk  obtained at 1.5" (right) for the several steps of the
    simulation. 
    }
  \label{fig:deconv_sb}
\end{figure*}

To evaluate the significance and the nature of the PSF patterns that may
appear after deconvolution of a coronagraphic image of a circumstellar
disk, we produce synthetic HST/ACS-like coronagraphic observations of
an edge-on disk that we afterwards deconvolve. Instead of using a Tiny
Tim PSF, we produce a synthetic HST PSF image using our own
diffraction code that we have developed to model the behavior and
performance of coronagraphs, as well as to gauge the potential of
several planet finding projects \citep{boccaletti04b, boccaletti05,
  boccaletti06}. 

We consider a $2.4$\,m diameter telescope observing at $0.606\,\mu$m,
with a bandwidth of $33$\%, and providing an angular resolution of
$0.052$" (similar to the ACS parameters).  We generate the image of an edge-on circumstellar disk
having a $4.5$'' diameter extension and a midplane surface brightness
proportional to $r^{-1.5}$, $r$ being the angular distance to the
star. The disk extends vertically over $0.52$" ($10$ times the
angular resolution) with a Gaussian profile of FWHM = $0.17$"
(Fig.\,\ref{fig:deconv_ima}, top panel).  The disk model is absolutely free of structures. 
Note that some of these parameters are arbitrary and are not intended to exactly model the \bp\ disk. 
This object contains $8542$
points for which we have to calculate the image through a coronagraph
(a Lyot mask of $1.8$" diameter) and for several wavelengths so as to
produce a realistic coronagraphic polychromatic image resembling that
of the ACS instrument. To simulate realistic images, the model
accounts for the phase aberrations using the measured phase map of the
HST \citep{krist95}, and we assume similar static aberrations on the
reference star but with an additional aberration in the form of a
defocus ($4\,$nm rms) to account for the so-called "PSF breathing"
(the HST PSF is slightly evolving over time). This amount of defocus is
taken arbitrarily as 
we are looking for qualitative results here. Another way to produce differences in the target star and
reference star images would be to apply a different pointing for the
two images with respect to the Lyot mask. The star to disk flux ratio
is also arbitrary and is selected to have the
residuals on the subtracted image of the order of the
disk intensity. We suspect that this is precisely in this regime of
contrast that deconvolution may generate artifacts. The synthetic
images is then deconvolved as in Sec.\,\ref{sec:warp}.

Several images to illustrate the process of image formation are
displayed in Fig.\,\ref{fig:deconv_ima}. 
 The two top panels show the disk model alone (no star) and the disk convolved with the PSF. The two lower panels are coronagraphic images (reference star subtracted) before and after deconvolution. Several patterns like speckles, rings, and radial spikes which were not included in the disk model are seen and are clearly enhanced in the deconvolution process. For instance, the radial spikes mimic the splitting of the disk midplane into two parts but are obviously not real. These spikes are aligned here with the disk to depict a worse case but could be differently oriented.

We used these simulated data to measure the averaged radial profile and the disk
thickness (Fig. \ref{fig:deconv_sb}), before and after deconvolution
of the image. Despite the intensity fluctuations resulting from the
PSF patterns the averaged radial profile is consistent with the
model except near the mask edges. It is interesting to see how the deconvolution
enhances the variations of the intensity profile
(Fig. \ref{fig:deconv_sb}, left) especially at distances closer than 1.0-1.2".
As a confirmation of the discussion in Sec.\,\ref{sec:mid} and contrary to what is proposed in
\citet{golimowski06}, the disk thickness is not affected by the
deconvolution as long as it is thicker than the angular resolution (Fig. \ref{fig:deconv_sb}, right). In
the simulation, the FWHM is only $0.17$" which is about 3 times the
angular resolution while in the actual data of \bp, the disk thickness
is about $0.87$" (Fig.\,\ref{fig:warp}) so the PSF blurring should be
even less visible. The vertical cut shown in Fig. \ref{fig:deconv_sb} (right) is obtained at 1.5" (left side of the image). The shape of the blue line (deconvolved image) changes significantly at closer separations, so the FWHM of the disk must be measured in a region less affected by PSF patterns.

Although the simple simulation presented here reveals the production
of artifacts, a more general analysis would be necessary to place
quantitative limitations when using deconvolution with coronagraphic
images of circumstellar disks. We note that the simulations presented here do not rule out the the reliability of the secondary disk shown in \citet{golimowski06}. This secondary disk is clearly significant with respect to PSF patterns arising in deconvolution of coronagraphic images. In particular, it is symmetrical and originates in the main disk at a distance where PSF patterns are much less preponderant (see Sec \ref {sec:warp}). The PSF patterns that may be confused with disks structures like spikes mostly originate from the star.

\section{Conclusion}
This paper reports on the observation of the \bp\ disk in the near IR with NACO at the VLT using two coronagraphs in two different spectral bands. We have used a careful data reduction process to get rid as much as possible of the diffraction residuals and artifacts that arise in high contrast imaging with a coronagraph on ground-based telescopes. The resulting images reveal the dusty disk in preferential regions depending on the type of coronagraph: between 1.2" to 7.5" with the Lyot mask and between 0.7" to 2.5" for the FQPM image. These two sets of data are therefore complementary in terms of spectral bands and angular separations. 

The Lyot mask coronagraph allows a very good characterization of the disk at large distances and especially of the famous warp.  We have been able to precisely measure the position, elevation and thickness of the warp and we have shown geometric and photometric particular asymmetries that could eventually been used as inputs to dynamical modeling in order to better constrain the characteristics of the forming planetary system. Some of these asymmetries were reported in various spectral bands (visible and mid IR). 
The inversion of the surface brightness profiles have confirmed the presence of an
asymmetric main planetesimal belt peaking at around $80$--$100$\,AU
responsible for the dust material beyond that distance. The inversion
further supports the presence of significant amounts of additional
material inside the main planetesimal belt as suggested by other
observations, depicting a complex planetary system.

The FQPM image reveals the presence of very close patterns very similar to what is observed in the mid-IR, a spectral regime that is sensitive to the dust emission. The interpretation of such structures would require further analysis possibly involving dynamical models to understand if the origin is similar to those observed in the mid-IR. 

To understand the issue of deconvolution in coronagraphic imaging we have performed numerical simulations to mimic HST Lyot coronagraphic data. Assuming a disk with no structure, this qualitative work shows how the data reduction process may emphasize some PSF structures that reinforce the need for a thorough analytical study for this type of data. 

These new images of the \bp\ debris disk presented here demonstrate the interest of ground-based instruments to achieve high angular resolution at longer wavelengths than with the HST and motivate for even more accurate observations. The sub-arcsec region is now receiving most of the attention with the recent discovery of a planet candidate. Constraints on the characteristics of this object and possibly others could be inferred form the spatial distribution of the dust.
Therefore, the precise registration of the inner structures we have performed if complemented with perturbation theory will help to better characterize this planetary system. To confirm the numerical simulations of \citet{Freistetter07} the study of the disk and its structures near 12\,AU is essential. This corresponds precisely to the limit reached in our data (0.62") and therefore the motivation is strong to perform more accurate observations. This is becoming possible at the VLT. The coronagraphic suite of NACO now implements other FQPMs tuned for the H and Ks bands with the ability to reach a better attenuation. Additional techniques  also have been installed (Angular Differential Imaging, Spectral Differential Imaging) that may facilitate the calibration of the residual starlight and hence the registration of structures even closer to the star. 

After more than 20 years of intensive studies, the dusty disk around \bp\ is still the subject of many questions. 
Direct imaging observations are complementary to spectroscopy and photometry and are mandatory to better constrain the spatial distribution of the dust, planetesimals and hypothetical planets in order to contribute to the understanding of the planetary formation phase.

\begin{acknowledgements}
We are grateful to the ESO staff supporting observations with NACO at the VLT and to the referee for an objective report that help to improve the manuscript. This work also received the support of PHASE, the high angular resolution partnership between ONERA, Observatoire de Paris, CNRS and University Denis Diderot Paris 7.
\end   {acknowledgements}

\bibliography{bpic_printer_v3}

\end{document}